\documentclass[journal]{IEEEtran}
\IEEEoverridecommandlockouts
\usepackage{amsmath,amssymb,amsfonts}
\usepackage{graphicx}
\usepackage{booktabs}
\usepackage{multirow}
\usepackage{array}
\usepackage[numbers, sort&compress]{natbib}
\usepackage[pagebackref=true, colorlinks=true, linkcolor=blue, citecolor=blue]{hyperref}
\usepackage[dvipsnames]{xcolor} 
\usepackage[utf8]{inputenc}
\usepackage{eso-pic}
\usepackage{tikz}
\usepackage{algorithm}
\usepackage{algpseudocode}
\usepackage{subcaption}
\usepackage{caption}
\usepackage{orcidlink}

\newcommand{\up}{\textcolor{ForestGreen}{\boldsymbol{\uparrow}}}
\newcommand{\down}{\textcolor{red}{\boldsymbol{\downarrow}}}

\usetikzlibrary{positioning,patterns,fit,arrows.meta,decorations.pathreplacing,calc}

\hypersetup{
    colorlinks=true,
    urlcolor=magenta,      
    linkcolor=blue,            
    citecolor=blue              
}

\begin{document}
\title{
\bfseries\LARGE
Do Transformers Actually Help Intrusion Detection?\\
 A Temporal Sequence Evaluation on CIC-IDS2017
}

\author{
\IEEEauthorblockN{
Zach Moczkodan$^{1}$\IEEEauthorrefmark{2}\,\orcidlink{0009-0003-6908-4935},~\IEEEmembership{Student Member,~IEEE},
Hany Ragab$^{2}$\IEEEauthorrefmark{2}\,\orcidlink{0000-0003-3167-9100},~\IEEEmembership{Member,~IEEE}
}

\thanks{
\IEEEauthorrefmark{2}Department of Electrical and Computer Engineering, Faculty of Engineering, Royal Military College of Canada (RMC), Kingston, Ontario, K7K 7B4, Canada.
Emails:
\href{mailto:zach.moczkodan@gmail.com}{\color{black}{\nolinkurl{zach.moczkodan@gmail.com}}}$^{1}$,
\href{mailto:zachery.moczkodan@forces.gc.ca}{\color{black}{\nolinkurl{zachery.moczkodan@forces.gc.ca}}}$^{1}$,
\href{mailto:hany.ragab@rmc-cmr.ca}{\color{black}{\nolinkurl{hany.ragab@rmc-cmr.ca}}}$^{2}$,
\href{mailto:hany.ragab@queensu.ca}{\color{black}{\nolinkurl{hany.ragab@queensu.ca}}}$^{2}$
}
}
\maketitle
\AddToShipoutPictureBG*{%
  \AtPageLowerLeft{%
    \raisebox{0.32in}{\hspace{0.68in}%
      \parbox{\dimexpr\paperwidth-1.37in\relax}{%
        \footnotesize ~\copyright~ 2026 IEEE. Personal use of this material is permitted. Permission from IEEE must be obtained for all other uses, in any current or future media, including reprinting/republishing this material for advertising or promotional purposes, creating new collective works, for resale or redistribution to servers or lists, or reuse of any copyrighted component of this work in other works.
      }%
    }%
  }%
}
\begin{abstract}
Recent deep learning approaches for network intrusion detection increasingly incorporate temporal architectures such as recurrent networks and Transformers, often reporting near-perfect performance on CIC-IDS2017. However, many existing studies neither supply their temporal modules with genuine sequence inputs nor evaluate under realistic, leakage-free conditions, making it unclear whether reported gains arise from true sequence-modelling capability.
In this work, we reformulate CIC-IDS2017 as a temporal intrusion-detection task by constructing ordered flow sequences from network conversations and benchmarking nine classical and deep learning architectures under a random split, two leakage-free splits, and a padding-scheme ablation.
The central finding is that \emph{padding convention, not architecture, determines the Transformer's performance}: on genuinely sequential (non-padded) windows the Transformer achieves the highest macro-F1 of any model in the experiment ($0.89$), yet under zero-pad+mask evaluation it drops markedly while LSTM, GRU, and 1D-CNN remain stable and a class-balanced Random Forest is the most robust model overall. The same repeat-last padding cue masks a $67\times$ increase in the Transformer's false-alarm rate that only leakage-free evaluation reveals.
Evaluation methodology --- specifically padding convention and split protocol --- thus has a larger effect on reported performance than architectural choice.
We advocate for leakage-free splits, explicit padding disclosure, and sequence-aware benchmarking as standard practice in future IDS research. Our code and implementation details are available here: \texttt{\url{https://github.com/zachmocz/temporal-ids-bench}}.
\end{abstract}


\section{Introduction}

\IEEEPARstart{I}{ntrusion} Detection Systems (IDS) are fundamental to the security of modern communication networks and cyber-physical systems~\cite{ring2019surveyids}. From connected vehicles and industrial control networks to critical infrastructure, safety-critical systems increasingly depend on real-time IDS to distinguish legitimate traffic from attacks under tight latency budgets.
With the increasing complexity of network traffic and the emergence of sophisticated attack strategies, machine learning--based IDS solutions have gained significant attention~\cite{hozouri2025surveydlids}. Among publicly available datasets, CIC-IDS2017~\cite{sharafaldin2018cicids} has become a widely adopted benchmark due to its realistic traffic generation and diversity of attack types.

Most existing approaches treat CIC-IDS2017 as a static tabular classification problem, where each network flow is independently classified. However, network traffic is inherently temporal, and ignoring sequential dependencies may limit detection performance, particularly for attacks whose behaviour emerges across multiple related flows. Recently, Transformer architectures~\cite{vaswani2017attention} have demonstrated strong performance on general time-series modelling tasks~\cite{zerveas2021transformerts,lim2021temporalfusion}, and recent Convolutional Neural Network (CNN)--Long Short-Term Memory (LSTM)--Transformer hybrids~\cite{liu2025cnnlstmtransformer,yao2023cnntransformer} report excellent accuracy on CIC-IDS2017.

\begin{figure}[!t]
\centering
\resizebox{\columnwidth}{!}{%
\begin{tikzpicture}[
    >=Stealth,
    font=\footnotesize,
    box/.style={
        rectangle, rounded corners=3pt, draw=black!55,
        align=center, minimum height=0.55cm,
        inner sep=3pt, text width=1.95cm
    },
    source/.style={box, fill=gray!8},
    stream/.style={
        box, fill=blue!8, draw=blue!60,
        text width=6.0cm, minimum height=0.62cm,
        font=\footnotesize\bfseries, text=blue!45!black
    },
    static/.style={
        box, fill=gray!8, text width=2.55cm,
        font=\footnotesize\bfseries
    },
    temporal/.style={
        box, fill=green!8, draw=green!45!black,
        text width=2.55cm,
        font=\footnotesize\bfseries, text=green!30!black
    },
    ids/.style={
        box, fill=violet!8, draw=violet!70,
        text width=2.35cm,
        font=\footnotesize\bfseries,
        text=violet!45!black
    },
    outstatic/.style={
        box, fill=gray!8, text width=2.45cm,
        font=\footnotesize\bfseries
    },
    outtemp/.style={
        box, fill=green!8, draw=green!45!black,
        text width=2.45cm,
        font=\footnotesize\bfseries,
        text=green!30!black
    },
    flowbox/.style={
        rectangle, rounded corners=2pt,
        minimum width=0.37cm,
        minimum height=0.23cm,
        inner sep=1pt,
        draw=black!15,
        fill=white
    },
    flowtemp/.style={
        flowbox,
        draw=green!55!black,
        fill=green!5,
        font=\scriptsize
    },
    flowstatic/.style={
        flowbox,
        draw=black!35,
        fill=white,
        font=\scriptsize
    },
    arrow/.style={->, line width=0.45pt, draw=black!35},
    faint/.style={line width=0.7pt, draw=black!45}
]

\node[source] (veh) at (-2.7,0) {Connected vehicles};
\node[source] (ind) at (0,0) {Industrial control};
\node[source] (inf) at (2.7,0) {Critical infrastructure};

\node[stream] (stream) at (0,-1.1) {Stream of network flows};

\draw[faint]
    (veh.south) -- ++(0,-0.28) -| ([xshift=-1.6cm]stream.north);

\draw[faint]
    (ind.south) -- ([yshift=0.02cm]stream.north);

\draw[faint]
    (inf.south) -- ++(0,-0.28) -| ([xshift=1.6cm]stream.north);

\node[static] (static) at (-1.95,-2.25) {Static model};
\node[temporal] (temporal) at (1.95,-2.25) {Temporal model};

\draw[arrow] (stream.south) -- ++(0,-0.35) -| (static.north);
\draw[arrow] (stream.south) -- ++(0,-0.35) -| (temporal.north);

\draw[dashed, draw=black!12] (0,-1.35) -- (0,-5.25);

\node[
    box,
    fill=gray!5,
    draw=black!15,
    text width=2.85cm,
    minimum height=0.42cm
] (sf) at (-1.95,-3.05) {};

\node[flowstatic] at (-2.75,-3.05) {$f_1$};
\node[flowbox]    at (-2.25,-3.05) {};
\node[flowbox]    at (-1.75,-3.05) {};
\node[flowbox]    at (-1.25,-3.05) {};
\node[flowbox]    at (-0.75,-3.05) {};

\node[font=\scriptsize, text=black!75] (sftxt) at (-1.95,-3.48)
    {one flow at a time};

\node[
    box,
    fill=green!4,
    draw=green!20,
    text width=2.85cm,
    minimum height=0.42cm
] (tf) at (1.95,-3.05) {};

\node[flowtemp] at (0.95,-3.05) {$f_1$};
\node[flowtemp] at (1.45,-3.05) {$f_2$};
\node[flowtemp] at (1.95,-3.05) {$f_3$};
\node[flowtemp] at (2.45,-3.05) {$f_4$};
\node[flowtemp] at (2.95,-3.05) {$f_5$};

\node[font=\scriptsize, text=black!75] (tftxt) at (1.95,-3.48)
    {per conversation};

\node[ids] (ids1) at (-1.95,-4.20) {IDS model};
\node[ids] (ids2) at (1.95,-4.20) {IDS model};

\draw[arrow] (static.south) -- (sf.north);
\draw[arrow] (sftxt.south) -- (ids1.north);

\draw[arrow] (temporal.south) -- (tf.north);
\draw[arrow] (tftxt.south) -- (ids2.north);

\node[outstatic] (out1) at (-1.95,-5.05) {\textbf{Normal / Attack}};
\node[outtemp]   (out2) at (1.95,-5.05) {\textbf{Normal / Attack}};

\draw[arrow] (ids1.south) -- (out1.north);
\draw[arrow] (ids2.south) -- (out2.north);

\end{tikzpicture}%
}
\caption{\textbf{Static versus temporal intrusion detection.} Static models classify one network flow at a time, whereas temporal models exploit ordered flow sequences within a conversation before producing the IDS decision.}
\label{fig:motivation}
\end{figure}
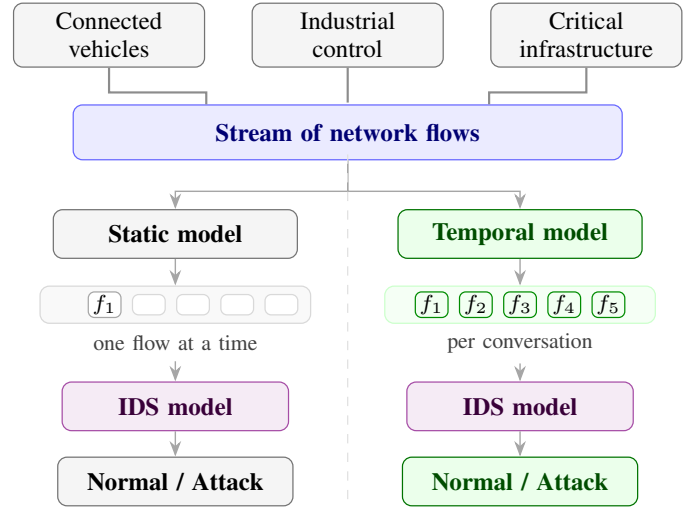

Despite the growing adoption of Transformer-based IDS architectures~\cite{kheddar2025surveyformer,manocchio2024flowtransformer,wu2022rtids}, it remains unclear whether their reported improvements arise from effective temporal modelling or from other aspects of the learning pipeline. Many existing studies evaluate these models under conventional random train/test splits and limited temporal formulations~\cite{liu2025cnnlstmtransformer,yao2023cnntransformer}, making it difficult to assess their robustness under realistic deployment conditions~\cite{engelen2021troubleshoot}.
This observation motivates the following research question:

\begin{quote}
\textit{Do Transformer-based architectures provide meaningful advantages for intrusion detection when network traffic is formulated as a genuine temporal sequence modelling problem and evaluated under realistic conditions?}
\end{quote}

To address this question, we reformulate CIC-IDS2017 as a temporal
sequence-classification task and benchmark nine architectures under
four evaluation conditions spanning split strategy, padding convention,
and sequence length. The main contributions are:
\begin{enumerate}
    \item A temporal evaluation framework for network intrusion detection
    on CIC-IDS2017, based on conversation-level flow sequences and
    leakage-aware, padding-explicit train/test protocols.

    \item A systematic benchmark of nine classical and deep learning
    architectures across a conventional random split, two leakage-free
    splits, and a padding-scheme ablation.

    \item An analysis of operational trade-offs, including false-alarm
    rate and inference latency, across all nine architectures and
    evaluation conditions.
\end{enumerate}

The paper is structured as follows: Section~\ref{sec:related} reviews related work on classical, sequential, and Transformer-based IDS approaches. Section~\ref{sec:methodology} describes the dataset, preprocessing pipeline, temporal windowing procedure, and models evaluated.
Section~\ref{sec:results} presents experimental results under the conventional random split, including sequence-length and padding ablations. Section~\ref{sec:realistic} evaluates all architectures under leakage-free protocols. Section~\ref{sec:discussion} discusses the findings, and Section~\ref{sec:conclusion} concludes the paper.

\section{Related Work}\label{sec:related}

\subsection{Classical and Sequential Baselines}

Traditional IDS approaches encompass Random Forests (RF)~\cite{breiman2001randomforest}, Support Vector Machines (SVMs)~\cite{cortes1995svm}, Multi-Layer Perceptrons (MLPs), and image-encoded CNNs that represent flow feature vectors as grayscale images; despite their simplicity, these methods remain competitive with deep learning when no genuine temporal structure is exploited~\cite{ring2019surveyids,hozouri2025surveydlids}. Sequential models, particularly LSTMs~\cite{hochreiter1997lstm} and Gated Recurrent Units (GRUs)~\cite{cho2014gru}, have since gained traction for their ability to capture temporal dependencies across related network flows~\cite{yin2017lstmids}.

\subsection{Transformers for Sequence Modelling and for Intrusion Detection}

Following Vaswani \emph{et al.}~\cite{vaswani2017attention}, the Transformer has been adapted to general time-series and sequence-representation learning. Zerveas \emph{et al.}~\cite{zerveas2021transformerts} introduce a Transformer framework for unsupervised multivariate time-series representation learning; Lim \emph{et al.}~\cite{lim2021temporalfusion} propose the Temporal Fusion Transformer for interpretable multi-horizon forecasting; K\"am\"ar\"ainen~\cite{kamarainen2025seqintro} provides an introductory treatment of the building blocks of Transformer sequence modelling. Kheddar~\cite{kheddar2025surveyformer} surveys the application of Transformers and large language models to intrusion detection specifically. In the IDS literature, the FlowTransformer framework~\cite{manocchio2024flowtransformer} provides configurable transformer blocks but reports that the classification head dominates performance, and does not compare against classical baselines; Wu \emph{et al.}~\cite{wu2022rtids} propose a robust Transformer-based IDS but evaluate only on the random split.

Liu \emph{et al.}~\cite{liu2025cnnlstmtransformer} propose a CNN--LSTM--Transformer hybrid that achieves 99.20\% test accuracy on CIC-IDS2017. Examination of their implementation, however, reveals that each flow
representation is reshaped to a sequence of length one before both the LSTM and Transformer stages, leaving neither temporal module with a temporal axis to leverage. The per-class sample counts reported in that work also differ substantially from the CIC-IDS2017 distribution (e.g., Heartbleed: 514 vs.\ 11 in the original; Infiltration: 6{,}401 vs.\ 36), indicating that their reported accuracy was not obtained on the standard dataset distribution.

Yao \emph{et al.}~\cite{yao2023cnntransformer} report 91.06\% accuracy (92.15\% under 10-fold cross-validation) on a similarly constrained formulation: their model operates on inputs of effective sequence length one and consolidates the 14 original attack categories into six, with Patator precision of $0.28$ and an Infiltration detection rate of $0.43$ on the resulting taxonomy. These limitations are not isolated to CIC-IDS2017; Transformer-based IDS classifiers applied to adjacent benchmarks show a consistent pattern of high reported accuracy achieved at the single-flow level, without recourse to genuine temporal sequence inputs.

\subsection{Why Temporal Modelling Matters---and What Is Missing}

A real IDS sees a continuous stream of flows; individual events become discriminative only in the context of recent activity from the same conversation. A port scan manifests as a burst of failed connection attempts; SSH-Patator as repeated login attempts on port~22; DoS-Slowloris as many simultaneous low-rate sessions held open. None of these patterns are visible in a single flow.
Three methodological gaps recur in recent CIC-IDS2017 papers~\cite{liu2025cnnlstmtransformer,yao2023cnntransformer}:
\textbf{(1)~Sequence length one}, so the temporal modules have no temporal axis to exploit;
\textbf{(2)~Padding artefacts}, where the choice of how to pad short conversations silently injects a detectable signal into self-attention; and
\textbf{(3)~Optimistic random splits}, where adjacent sliding windows from the same five-tuple conversation can land on opposite sides of the train/test boundary, inflating apparent generalisation by exposing models to conversation-specific patterns at test time.
Engelen \emph{et al.}~\cite{engelen2021troubleshoot} additionally caution that CIC-IDS2017 contains label noise and flow-construction artefacts that can further inflate reported accuracy.

\section{Methodology}\label{sec:methodology}

\begin{figure*}[!t]
\centering
\begin{tikzpicture}[
    >=Stealth,
    font=\footnotesize,
    stage/.style={
        rectangle, draw=black!70, rounded corners=2.5pt, align=center,
        minimum height=0.85cm, inner sep=3pt, fill=blue!5,
        text width=2.35cm, font=\footnotesize,
    },
    modelblock/.style={
        rectangle, draw=black!70, rounded corners=2.5pt, align=center,
        minimum height=0.85cm, inner sep=3pt, fill=orange!12,
        text width=2.55cm, font=\footnotesize,
    },
    evalblock/.style={
        rectangle, draw=black!70, rounded corners=2.5pt, align=center,
        minimum height=0.85cm, inner sep=3pt, fill=green!10,
        text width=2.95cm, font=\footnotesize,
    },
    arrow/.style={->, line width=0.6pt, >=Stealth},
    sub/.style={font=\scriptsize, color=black!70},
]

\node[stage] (s1) {\textbf{CIC-IDS2017}\\\textcolor{black!70}{\scriptsize $2.83$M flows, $15$ classes}};
\node[stage, right=0.32cm of s1] (s2) {\textbf{Group by 5-tuple}\\\textcolor{black!70}{\scriptsize sort by timestamp}};
\node[stage, right=0.32cm of s2] (s3) {\textbf{Sliding} $T{=}20$ \textbf{windows}\\\textcolor{black!70}{\scriptsize $630{,}264$ windows}};
\node[modelblock, right=0.32cm of s3] (s4) {\textbf{CNN + Transformer}\\\textcolor{black!70}{\scriptsize PE + [CLS] + masked attn.}};
\node[evalblock, right=0.32cm of s4] (s5) {\textbf{$9$ models, $4$ protocols}\\\textcolor{black!70}{\scriptsize random / time / group / no-pad}};

\draw[arrow] (s1) -- (s2);
\draw[arrow] (s2) -- (s3);
\draw[arrow] (s3) -- (s4);
\draw[arrow] (s4) -- (s5);

\end{tikzpicture}
\caption{\textbf{Main contribution.} We re-formulate CIC-IDS2017 as a real temporal sequence task by grouping flows on their five-tuple and constructing $T{=}20$ sliding windows, then benchmark nine architectures across four evaluation protocols.}
\label{fig:contribution}
\end{figure*}
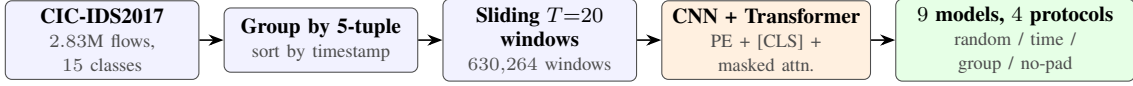

Fig.~\ref{fig:contribution} summarises the pipeline at a glance. Static baselines branch off after the windowing stage and consume only the last flow of each window; the CNN--Transformer adds sinusoidal positional encoding, prepends a learnable [CLS] token, runs the result through two masked Transformer encoder blocks, and classifies from the [CLS] state.

\subsection{Dataset and Preprocessing Pipeline}

We use the full CIC-IDS2017 capture~\cite{sharafaldin2018cicids}: $2{,}827{,}677$ labelled flows over five days, comprising one benign class and fourteen attack categories. Table~\ref{tab:dropped} lists every column we removed from the model feature set and why; $76$ numeric flow features remain after dropping.

\begin{table}[!t]
\centering
\caption{Preprocessing: columns removed from the model feature set before training. Every dropped column is either an identifier, a timestamp, or the label itself.}
\label{tab:dropped}
\renewcommand{\arraystretch}{1.05}
\footnotesize
\setlength{\tabcolsep}{6pt}
\begin{tabular}{@{}p{2.6cm}p{0.8cm}p{4.7cm}@{}}
\toprule
\textbf{Column} & \textbf{Type} & \textbf{Reason for removal} \\
\midrule
\texttt{flow\_id}        & Identifier  & Per-flow unique key; would memorise. \\
\texttt{source\_ip}      & 5-tuple     & Used for grouping only. \\
\texttt{destination\_ip} & 5-tuple     & Used for grouping only. \\
\texttt{source\_port}    & 5-tuple     & Used for grouping only. \\
\texttt{destination\_port} & 5-tuple   & Used for grouping only.\\
\texttt{protocol}        & 5-tuple     & Used for grouping only. \\
\texttt{Timestamp}       & Time        & Used for ordering and time-split only. \\
\texttt{attack\_label}   & Label       & Target. \\
\midrule
\multicolumn{3}{@{}l}{\textbf{Retained:} 76 numeric flow features (CIC-FlowMeter output).} \\
\bottomrule
\end{tabular}
\end{table}

The pipeline proceeds as follows: (1)~replace $\pm\infty$ with NaN and remove the affected rows; (2)~encode the 15-class label; (3)~sort by \texttt{Timestamp}; (4)~group by five-tuple and build $T{=}20$ sliding windows; (5)~exclude windowed classes with fewer than two examples (Heartbleed, DoS Slowhttptest), yielding a 13-class subset that covers $\geq 99.6\%$ of flows; (6)~apply a stratified 80/20 split by label; (7)~apply $z$-score standardisation fitted on training data only, plus min-max scaling for the image encoders; (8)~apply inverse-frequency class weights $w_c = N/(13 \cdot N_c)$ for the deep models, where $N$ is the total number of training windows and $N_c$ is the number of windows in class $c$. We do not perform low-variance pruning or principal component analysis (PCA).

\subsection{Five-Tuple Temporal Windowing}

Let $\mathcal{D} = \{(\mathbf{x}_i, y_i, t_i, q_i)\}_{i=1}^{N}$ denote the dataset of $N$ network flows, where $\mathbf{x}_i \in \mathbb{R}^{F}$ is a flow feature vector with $F=76$ numeric attributes, $y_i \in \mathcal{Y}$ is the attack label drawn from the label set $\mathcal{Y}$ (with $|\mathcal{Y}|{=}13$), $t_i$ is the timestamp, and $q_i = (\text{src ip}, \text{dst ip}, \text{src port}, \text{dst port}, \text{protocol})$ is the canonical five-tuple. We partition $\mathcal{D}$ into per-conversation buckets $\mathcal{G}_q = \{ i : q_i = q \}$, sort each bucket by timestamp, and take sliding windows of length $T$:
\begin{equation}
\mathbf{X}_{q,s} = \bigl[\mathbf{x}_{i_s}, \mathbf{x}_{i_{s+1}}, \ldots, \mathbf{x}_{i_{s+T-1}}\bigr] \in \mathbb{R}^{T \times F}.
\end{equation}
The label of each window is the class of its final flow $y_{q,s} = y_{i_{s+T-1}}$, mimicking how a streaming IDS would classify in real time. Conversations of length~$1$ are skipped. Conversations shorter than $T$ are completed by repeating the last real flow---the conventional padding scheme used in prior IDS work~\cite{liu2025cnnlstmtransformer,yao2023cnntransformer}---which we adopt as the primary protocol so the random-split numbers are directly comparable. Section~\ref{sec:padding} re-runs the same models with zero-padding plus an explicit attention mask. To prevent a single very long DDoS conversation from dominating training we cap any one five-tuple at $K{=}50$ windows.

\begin{figure}[!t]
\centering
\hspace{-0.4cm}
\begin{tikzpicture}[
    >=Stealth,
    font=\footnotesize,
    real/.style={
        rectangle, draw=green!50!black, fill=green!15,
        minimum width=0.52cm, minimum height=0.52cm,
        inner sep=0pt, align=center
    },
    pad_repeat/.style={
        rectangle, draw=gray!55, fill=gray!20,
        minimum width=0.52cm, minimum height=0.52cm,
        inner sep=0pt, align=center,
        pattern=north east lines, pattern color=gray!50
    },
    pad_zero/.style={
        rectangle, draw=red!50, fill=red!6,
        minimum width=0.52cm, minimum height=0.52cm,
        inner sep=0pt, align=center
    },
    mask1/.style={
        rectangle, draw=black!40, fill=black!8,
        minimum width=0.52cm, minimum height=0.30cm,
        inner sep=0pt, align=center, font=\scriptsize
    },
    mask0/.style={
        rectangle, draw=red!40, fill=red!8,
        minimum width=0.52cm, minimum height=0.30cm,
        inner sep=0pt, align=center, font=\scriptsize
    },
    dots/.style={font=\normalsize, text=black!55},
    lbl/.style={font=\footnotesize\bfseries, text=black},
    sublbl/.style={font=\scriptsize, text=black!65},
    brace/.style={decorate,
        decoration={brace, amplitude=4pt, mirror}, thick, draw=black!50},
    ann/.style={font=\scriptsize, text=black!55, anchor=west}
]

\def\rowA{0.00}
\def\rowB{-1.70}
\def\rowC{-3.25}

\def\xa{0}
\def\xb{0.62}
\def\xc{1.24}
\def\xd{1.86}
\def\xe{2.48}
\def\xf{3.10}
\def\xg{3.72}
\def\xh{4.34}

\node[lbl, anchor=east] at (-0.30, \rowA+0.26) {(1) Repeat-last};
\node[sublbl, anchor=east] at (-0.30, \rowA-0.05) {(primary)};
\node[lbl, anchor=east] at (-0.30, \rowB+0.26) {(2) Zero-pad\,+\,mask};
\node[sublbl, anchor=east] at (-0.30, \rowB-0.05) {(ablation)};
\node[lbl, anchor=east] at (-0.30, \rowC+0.26) {(3) Non-padded};
\node[sublbl, anchor=east] at (-0.30, \rowC-0.05) {(filter)};

\node[real] (A1) at (\xa,\rowA) {\scriptsize $f_{1}$};
\node[real] (A2) at (\xb,\rowA) {\scriptsize $f_{2}$};
\node[real] (A3) at (\xc,\rowA) {\scriptsize $f_{3}$};
\node[dots]      at (\xd,\rowA) {$\cdots$};
\node[real] (A5) at (\xe,\rowA) {\scriptsize $f_{k}$};
\node[pad_repeat] (A6) at (\xf,\rowA) {\scriptsize $f_{k}$};
\node[dots]      at (\xg,\rowA) {$\cdots$};
\node[pad_repeat] (A8) at (\xh,\rowA) {\scriptsize $f_{k}$};

\draw[brace] ($(A1.south west)+(0,-0.06)$) -- ($(A5.south east)+(0,-0.06)$)
    node[midway, below=3pt, font=\scriptsize, text=green!40!black] {$k$ real flows};
\draw[brace] ($(A6.south west)+(0,-0.06)$) -- ($(A8.south east)+(0,-0.06)$)
    node[midway, below=3pt, font=\scriptsize, text=gray!60!black] {$f_{k}$ repeated};

\node[real] (B1) at (\xa,\rowB) {\scriptsize $f_{1}$};
\node[real] (B2) at (\xb,\rowB) {\scriptsize $f_{2}$};
\node[real] (B3) at (\xc,\rowB) {\scriptsize $f_{3}$};
\node[dots]      at (\xd,\rowB) {$\cdots$};
\node[real] (B5) at (\xe,\rowB) {\scriptsize $f_{k}$};
\node[pad_zero] (B6) at (\xf,\rowB) {\scriptsize $\mathbf{0}$};
\node[dots]      at (\xg,\rowB) {$\cdots$};
\node[pad_zero] (B8) at (\xh,\rowB) {\scriptsize $\mathbf{0}$};

\node[mask1] at (\xa,\rowB-0.50) {\textbf{1}};
\node[mask1] at (\xb,\rowB-0.50) {\textbf{1}};
\node[mask1] at (\xc,\rowB-0.50) {\textbf{1}};
\node[dots, font=\footnotesize] at (\xd,\rowB-0.50) {$\cdots$};
\node[mask1] at (\xe,\rowB-0.50) {\textbf{1}};
\node[mask0] at (\xf,\rowB-0.50) {\textbf{0}};
\node[dots, font=\footnotesize] at (\xg,\rowB-0.50) {$\cdots$};
\node[mask0] at (\xh,\rowB-0.50) {\textbf{0}};
\node[sublbl, anchor=east] at (-0.30, \rowB-0.50) {\textit{mask}};

\node[real] (C1) at (\xa,\rowC) {\scriptsize $f_{1}$};
\node[real] (C2) at (\xb,\rowC) {\scriptsize $f_{2}$};
\node[real] (C3) at (\xc,\rowC) {\scriptsize $f_{3}$};
\node[dots]      at (\xd,\rowC) {$\cdots$};
\node[real] (C5) at (\xe,\rowC) {\scriptsize $f_{k}$};
\node[real] (C6) at (\xf,\rowC) {\scriptsize $f_{k+1}$};
\node[dots]      at (\xg,\rowC) {$\cdots$};
\node[real] (C8) at (\xh,\rowC) {\scriptsize $f_{T}$};

\draw[brace] ($(C1.south west)+(0,-0.06)$) -- ($(C8.south east)+(0,-0.06)$)
    node[midway, below=3pt, font=\scriptsize, text=green!40!black]
    {all $T$ positions are real flows};

\draw[<->, draw=black!45, line width=0.5pt]
    ($(A1.north west)+(0,0.20)$) -- ($(A8.north east)+(0,0.20)$)
    node[midway, above=1pt, font=\scriptsize, text=black!55] {window length $T$};

\draw[dashed, draw=black!25, line width=0.6pt]
    ($(A5.north east)+(0.02,0.16)$) -- ($(B5.south east)+(0.02,-0.64)$);

\node[real,       label={[font=\scriptsize]right:real flow}]     at (0.15, \rowC-1.20) {};
\node[pad_repeat, label={[font=\scriptsize]right:repeated flow}] at (1.84, \rowC-1.20) {};
\node[pad_zero,   label={[font=\scriptsize]right:zero vector}]   at (3.95, \rowC-1.20) {};

\end{tikzpicture}
\caption{Padding conditions for a conversation of $k$ real flows in a window of length $T$ ($k < T$). \emph{Repeat-last} : copies the final flow $f_{k}$ into remaining positions, creating an exploitable repetition signal. \emph{Zero-pad\,+\,mask} (ablation, Section~\ref{sec:padding}): pads with zero vectors and explicitly masks them from attention (LSTM/GRU use a Keras \texttt{Masking} layer). \emph{Non-padded subset} (filter, Section~\ref{sec:nonpad}): evaluates only windows with $T$ genuine flows, eliminating padding artefacts.}
\label{fig:padding_schemes}
\end{figure}
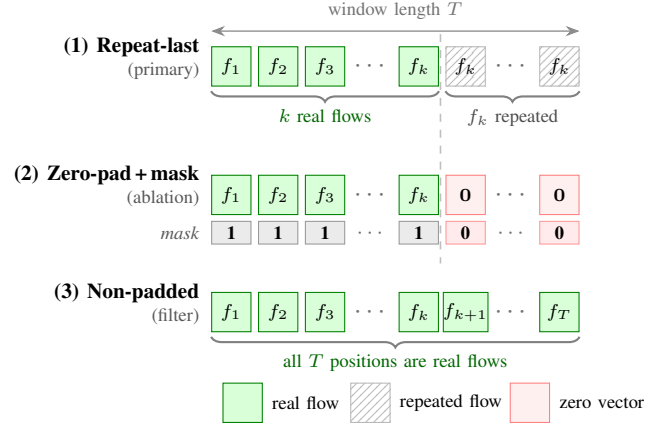

After windowing, ${\sim}1.46$\,M five-tuple groups yield approximately $630\,$k windows ($\sim$73\% padded), with the 13-class final set splitting 80/20 into ${\sim}504\,$k training and ${\sim}126\,$k test windows for the random-split protocol. The leakage-free splits (Section~\ref{sec:realistic}) produce slightly different totals because windows are constructed after the train/test partition.


\subsection{Train/Test Splits}
\label{sec:splits}

We evaluate every model under three split protocols. The \textbf{random} split is the convention used in prior CNN--LSTM--Transformer IDS papers; we adopt it as our baseline protocol to enable direct comparison with previously reported results. The \textbf{time-based} split sorts all flows by timestamp and uses the first 80\% for training, last 20\% for test---this answers ``can the model generalise from earlier traffic to later traffic?''. The \textbf{group-by-five-tuple} split partitions five-tuples 80/20 so every window from a given conversation lands on one side---this answers ``can the model generalise to unseen conversations?''. For both leakage-free splits, windows are constructed \emph{after} the split so no window straddles the train/test boundary.

\textbf{Padding protocol for leakage-free splits.} Both leakage-free splits use zero-pad+mask windows (as illustrated in Fig.~\ref{fig:padding_schemes}), so their results reflect a simultaneous change in both split strategy and padding convention. Section~\ref{sec:padding} isolates the padding contribution by holding the split fixed; Section~\ref{sec:realistic} holds the padding fixed and varies the split. Under zero-padding, static models extract the last \emph{real} flow from each window using the mask-sum position rather than the final array slot, which would otherwise be all zeros for padded windows.

\subsection{Models Evaluated}

We evaluate three categories of models, with all hyperparameters detailed in Table~\ref{tab:hyper}. 

\begin{table*}[!t]
\centering
\caption{Architecture and training hyperparameters. Static (non-temporal) models
consume only the last flow; temporal models consume the full $T{=}20$ window.
``Subsample'' indicates a class-stratified sub-sample of the training set;
``Full'' indicates the complete $504{,}211$-window training set.}
\label{tab:hyper}
\renewcommand{\arraystretch}{1.05}
\footnotesize
\setlength{\tabcolsep}{3pt}
\begin{tabular*}{\textwidth}{@{\extracolsep{\fill}}p{2.4cm}lcccl@{}}
\toprule
\textbf{Model} & \textbf{Architecture} & \textbf{Params} & \textbf{Batch} & \textbf{LR} & \textbf{Training set} \\
\midrule
Linear SVM        & \texttt{LinearSVC}, $C{=}1.0$, balanced class weights, 2000 max iters   & ---    & ---  & ---                  & Subsample, ${\sim}$108k$^{\dagger}$ \\
Random Forest     & 100 trees, max-depth 20, balanced class weights                          & ---    & ---  & ---                  & Subsample, ${\sim}$108k$^{\dagger}$ \\
MLP               & Dense(128) -- Dense(64) -- Dense(13, softmax)                            & 18.9k  & 1024 & $1\!\times\!10^{-3}$ & Full \\
Image-CNN         & Conv2D(32)--BN--MP--Conv2D(64)--BN--MP--Dense(128)--Drop(0.5)            & 53.6k  & 1024 & $1\!\times\!10^{-3}$ & Full \\
LSTM              & LSTM(64) -- LSTM(32) -- Dense(64) -- Drop(0.3) -- Dense(13)              & 51.9k  & 1024 & $1\!\times\!10^{-3}$ & Full \\
GRU               & GRU(64) -- GRU(32) -- Dense(64) -- Drop(0.3) -- Dense(13)               & 39.6k  & 1024 & $5\!\times\!10^{-4}$ & Full \\
1D-CNN            & Conv1D(64)--BN--MP--Conv1D(128)--BN--GAP--Dense(64)--Drop(0.3)          & 48.8k  & 1024 & $1\!\times\!10^{-3}$ & Full \\
Transformer       & TimeDist(CNN-128)--PE--[CLS]--$2{\times}$Block($d{=}128,h{=}4,d_{\text{ff}}{=}256$) & 318.7k & 256 & $5\!\times\!10^{-4}$ & Full \\
Transformer-small & TimeDist(CNN-64)--PE--[CLS]--$1{\times}$Block($d{=}64,h{=}2,d_{\text{ff}}{=}128$)  & 69.8k  & 256  & $5\!\times\!10^{-4}$ & Full \\
\bottomrule
\multicolumn{6}{l}{\footnotesize All deep models: Adam~\cite{kingma2015adam}, grad-clip norm 1.0, sparse CE loss, early-stop patience 6 (val-loss), ReduceLROnPlateau (factor 0.5, patience 2, floor $10^{-6}$),} \\
\multicolumn{6}{l}{\footnotesize $\leq$30 epochs, val-split 0.15, inv-freq class weights, seeds $\{2,42,123\}$. $^{\dagger}$SVM/RF: per-class cap of 50{,}000 windows, ${\approx}$108k total.} \\ 
\end{tabular*}
\end{table*}

\noindent The architectures are summarized as follows:

\textbf{1) Static (non-temporal) models}, consuming only the last flow of each window:
\begin{itemize}
    \item \textbf{Linear SVM}~\cite{cortes1995svm}: \texttt{LinearSVC} with regularisation strength $C{=}1.0$, balanced class weights.
    \item \textbf{Random Forest}~\cite{breiman2001randomforest}: 100 trees, max-depth 20, balanced class weights.
    \item \textbf{Multi-Layer Perceptron}~\cite{goodfellow2016dl}: Dense(128)--Dense(64)--Dense(13, softmax).
    \item \textbf{Image-CNN}: two stacked Conv2D--BN--MaxPool blocks (32 then 64 filters) applied to a $9{\times}9$ grayscale image encoding of the 76 flow features, followed by Dense(128) and Dropout(0.5)~\cite{ioffe2015batchnorm,srivastava2014dropout}.
\end{itemize}

\textbf{2) Sequential (temporal) models}, consuming the full \mbox{$(T \times F)$} window:
\begin{itemize}
    \item \textbf{LSTM}~\cite{hochreiter1997lstm}: stacked $64{\to}32$ unidirectional cells.
    \item \textbf{GRU}~\cite{cho2014gru}: identical topology with GRU cells.
    \item \textbf{1D-CNN over time}: $\text{Conv1D}_{64}{-}\text{Conv1D}_{128}$ with batch normalisation and global average pooling.
\end{itemize}

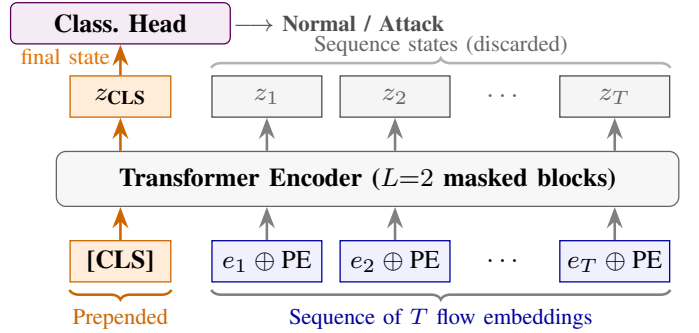
\begin{figure}[!t]
\centering
\resizebox{\columnwidth}{!}{%
\begin{tikzpicture}[
    >=Stealth,
    font=\footnotesize,
    emb/.style={
        rectangle, draw=blue!60!black, fill=blue!6,
        minimum width=1.2cm, minimum height=0.45cm,
        inner sep=0pt, align=center
    },
    cls/.style={
        rectangle, draw=orange!80!black, fill=orange!15,
        minimum width=1.1cm, minimum height=0.45cm,
        inner sep=0pt, align=center, font=\footnotesize\bfseries
    },
    outblk/.style={
        rectangle, draw=gray!60!black, fill=gray!8,
        minimum width=1.2cm, minimum height=0.45cm,
        inner sep=0pt, align=center, text=gray!80!black
    },
    head/.style={
        rectangle, draw=violet!70!black, fill=violet!8,
        rounded corners=2pt, minimum width=2.4cm, minimum height=0.45cm,
        align=center, font=\footnotesize\bfseries
    },
    trans/.style={
        rectangle, draw=black!60, fill=gray!6,
        rounded corners=3pt, minimum width=6.8cm, minimum height=0.6cm,
        align=center, font=\footnotesize\bfseries
    },
    arrow/.style={->, draw=black!50, thick},
    brace/.style={decorate, decoration={brace, amplitude=3pt, mirror}, thick, draw=black!50}
]

\def\xc{0}
\def\xa{1.6}
\def\xb{3.0}
\def\xd{4.2}
\def\xT{5.4}

\def\yin{0}
\def\ytr{0.9}
\def\yout{1.8}
\def\yhead{2.6}

\node[cls] (inC) at (\xc, \yin) {[CLS]};
\node[emb] (in1) at (\xa, \yin) {$e_1 \oplus \text{PE}$};
\node[emb] (in2) at (\xb, \yin) {$e_2 \oplus \text{PE}$};
\node      (ind) at (\xd, \yin) {$\cdots$};
\node[emb] (inT) at (\xT, \yin) {$e_T \oplus \text{PE}$};

\node[trans] (tr) at (2.7, \ytr) {Transformer Encoder ($L{=}2$ masked blocks)};

\node[cls] (ouC) at (\xc, \yout) {$z_{\text{CLS}}$};
\node[outblk] (ou1) at (\xa, \yout) {$z_1$};
\node[outblk] (ou2) at (\xb, \yout) {$z_2$};
\node[text=gray!80!black] (oud) at (\xd, \yout) {$\cdots$};
\node[outblk] (ouT) at (\xT, \yout) {$z_T$};

\node[head] (head) at (\xc, \yhead) {Class. Head};
\node[anchor=west, font=\scriptsize, text=black!70, xshift = -0.1cm] at (head.east) {$\longrightarrow$ \textbf{Normal / Attack}};

\draw[arrow, draw=orange!80!black] (inC) -- (inC |- tr.south);
\draw[arrow] (in1) -- (in1 |- tr.south);
\draw[arrow] (in2) -- (in2 |- tr.south);
\draw[arrow] (inT) -- (inT |- tr.south);

\draw[arrow, draw=orange!80!black] (ouC |- tr.north) -- (ouC);
\draw[arrow] (ou1 |- tr.north) -- (ou1);
\draw[arrow] (ou2 |- tr.north) -- (ou2);
\draw[arrow] (ouT |- tr.north) -- (ouT);

\draw[arrow, draw=orange!80!black] (ouC) -- (head)
    node[midway, right, font=\scriptsize, text=orange!90!black, yshift=0.04cm, xshift=-1.22cm] {final state};

\draw[brace] ($(in1.south west)+(0,-0.08)$) -- ($(inT.south east)+(0,-0.08)$)
    node[midway, below=2pt, font=\scriptsize, text=blue!60!black] {Sequence of $T$ flow embeddings};

\draw[brace, draw=orange!80!black] ($(inC.south west)+(0,-0.08)$) -- ($(inC.south east)+(0,-0.08)$)
    node[midway, below=2pt, font=\scriptsize, text=orange!80!black] {Prepended};

\draw[brace, decoration={mirror=false}, draw=gray!60] ($(ou1.north west)+(0,0.08)$) -- ($(ouT.north east)+(0,0.08)$)
    node[midway, above=0pt, font=\scriptsize, text=gray!80!black] {Sequence states (discarded)};

\end{tikzpicture}%
}
\caption{Information flow through the CNN--Transformer architecture. A learnable \texttt{[CLS]} token is prepended to the position-encoded flow embeddings ($e_t$). After self-attention processing, only the final state of the \texttt{[CLS]} token ($z_{\text{CLS}}$) is passed to the classification head, aggregating the entire sequence context into a single prediction.}
\label{fig:cls_token}
\end{figure}

\textbf{3) CNN--Transformer encoder} (our principal model). Each flow vector is min-max scaled, zero-padded from $76$ to $81$ values, and reshaped into a $9\times9$ grayscale image. A Conv2D feature extractor wrapped in a \texttt{TimeDistributed} layer produces an embedding sequence
\begin{equation}
\mathbf{e}_t = \text{CNN}(\mathbf{I}_t) \in \mathbb{R}^{D},\qquad t=1,\ldots,T,
\end{equation}
with $D{=}128$. Sinusoidal positional encoding~\cite{vaswani2017attention} is added,
\begin{align}
\text{PE}(t,2j)   &= \sin\!\left(\frac{t}{10000^{2j/D}}\right), \\
\text{PE}(t,2j{+}1) &= \cos\!\left(\frac{t}{10000^{2j/D}}\right),
\end{align}
and a learnable [CLS] token~\cite{devlin2019bert} is prepended (sequence length becomes $T{+}1$). The result is processed by $L{=}2$ encoder blocks, each performing multi-head self-attention with $h{=}4$ heads, where
$\mathbf{Q}$, $\mathbf{K}$, and $\mathbf{V}$ are the query, key, and value
matrices and $d_k = D/h$ is the per-head key dimension,
\begin{equation}
\mathrm{Attention}(\mathbf{Q},\mathbf{K},\mathbf{V}) =
\mathrm{softmax}\!\left(\frac{\mathbf{Q}\mathbf{K}^\top}{\sqrt{d_k}}\right)\mathbf{V},
\end{equation}
followed by a feed-forward sub-network of width $d_{\text{ff}}{=}256$, with residual connections and layer normalisation~\cite{ba2016layernorm}. The classification head is the [CLS] token's state at the output of the final block. Whenever a window contains zero-padded positions (Section~\ref{sec:padding}, Section~\ref{sec:realistic}), an explicit attention mask is passed into every \texttt{MultiHeadAttention} call, and the padded positions are re-zeroed immediately after the positional encoding is added (defence in depth). Fig.~\ref{fig:cls_token} illustrates the information flow, highlighting the prepended [CLS] token and its final state extraction.

A second variant, \textbf{Transformer-small}, tests whether the headline Transformer is over-parameterised relative to the available temporal signal: 1 encoder block, 2 heads, $D{=}64$, $d_{\text{ff}}{=}128$, $\sim$70k parameters versus $\sim$320k for the full Transformer.

\subsection{Evaluation Metrics}

The dataset is heavily class-imbalanced (BENIGN dominates at $80.3\%$ of flows, $88.5\%$ of windows), so we treat \textbf{macro-F1} as the headline metric: it weights every class equally regardless of frequency. Weighted-F1 saturates near $1.0$ for almost every model because of BENIGN's dominance, so we report it but do not rely on it. We also report macro precision, macro recall, ROC-AUC and PR-AUC (preferred under heavy imbalance~\cite{saito2015prroc}), together with the operational false-alarm rate (FPR: the fraction of BENIGN windows flagged as any attack class) and missed-attack rate (FNR: the fraction of attack windows predicted BENIGN). Every experiment is run with three random seeds ($\{2, 42, 123\}$); we report mean and standard deviation across seeds, with the seed controlling weight initialisation, training-set shuffling and (for the group-by-five-tuple split) the partition itself. Given the small number of seeds, these intervals are indicative rather than inferential; throughout, we treat differences within one per-seed standard deviation as statistical ties rather than rankings. Inference latency was evaluated using direct \texttt{model(x)} calls on an NVIDIA A100 GPU (40\,GB memory). Reported values represent the average of twenty runs on a batch of 128 windows for each $(T,\text{model})$ configuration. All experiments were performed using TensorFlow~2.20 on a system with 80\,GB RAM.

\section{Experimental Results}\label{sec:results}

\subsection{Overall Performance}

\begin{figure}[!t]
\centering
\includegraphics[width=\columnwidth]{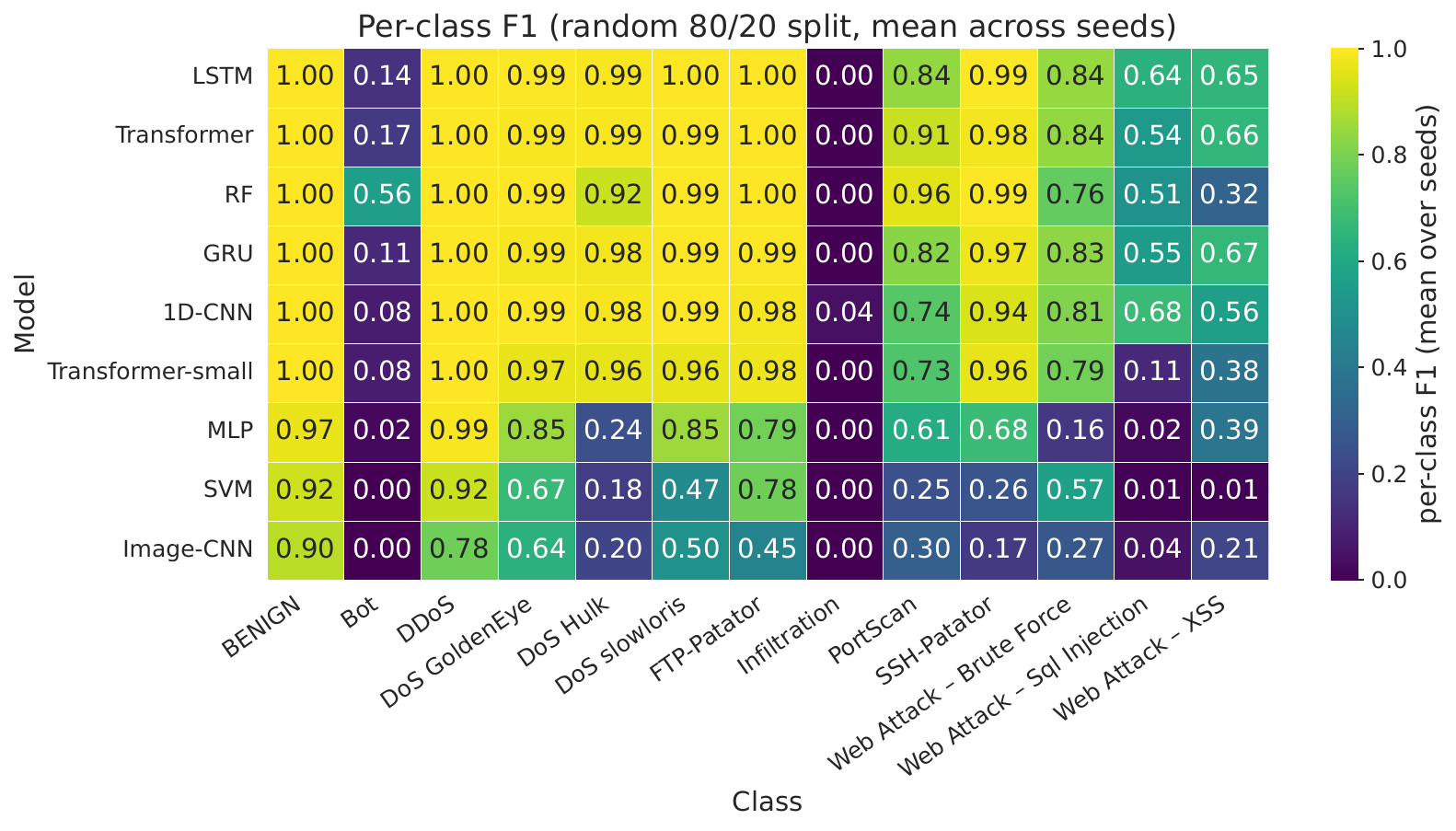}
\caption{Per-class F1 under the random 80/20 split, mean across three seeds. BENIGN and the high-volume DoS classes are detected by every temporal model at F1 ${\approx}\,1.0$; Infiltration is essentially undetectable across every architecture (4 training windows); Bot, Web Attack--SQL Injection and Web Attack--XSS show the largest spread among the leading models.}
\label{fig:perclass-f1-random}
\end{figure}

Table~\ref{tab:ids_splits} reports every model on the random 80/20 split, mean${}\pm{}$std across three seeds. The LSTM attains the highest macro-F1 ($0.776$), statistically indistinguishable from the CNN--Transformer ($0.775$), followed by Random Forest ($0.769$), GRU ($0.762$), 1D-CNN ($0.754$) and Transformer-small ($0.685$). The five leading models cluster within $0.022$ macro-F1; the LSTM's lead over RF is $0.007$, and over the Transformer just $0.001$, both well within the per-seed standard deviation of either model. Even on the protocol most favourable to it, the Transformer only ties the LSTM, and the temporal models do not separate cleanly from a class-balanced Random Forest. The remaining static models (SVM, MLP, image-CNN) trail substantially in macro-F1 despite high accuracy, reflecting their failure to detect minority classes from a single flow. On the rare classes the spread is wider (per-class breakdown in Fig.~\ref{fig:perclass-f1-random}): Random Forest leads on Bot (F1 $0.56$, mean across seeds), the temporal models span $0.38$--$0.67$ on Web Attack--XSS, and Infiltration is essentially undetectable across every model (best mean F1 $0.04$). Section~\ref{sec:realistic} re-evaluates the same nine models under the leakage-free group-by-five-tuple split.

\begin{table*}[!t]
\centering
\caption{Performance comparison on CIC-IDS2017 under the random 80/20 split, mean${}\pm{}$std over three seeds ($\{2, 42, 123\}$). Static models observe only the last flow of each window; temporal models process the full $T{=}20$ window. Best macro-F1 in \textbf{bold}; rows ordered within each block by macro-F1.}
\label{tab:ids_splits}
\renewcommand{\arraystretch}{1.05}
\footnotesize
\setlength{\tabcolsep}{4pt}
\begin{tabular}{lccccccc}
\toprule
\textbf{Model} & \textbf{Precision} & \textbf{Recall} & $\mathbf{F_1\text{-w}}$ & \textbf{ROC-AUC} & \textbf{PR-AUC} & \textbf{Macro $F_1$} & \textbf{Train (s)} \\
\midrule
\multicolumn{8}{l}{\textit{Static models}} \\
RF             & $0.77 \pm .01$ & $0.78 \pm .01$ & $0.998$        & $0.99 \pm .02$ & $0.79 \pm .01$ & $0.77 \pm .01$ & 4 \\
MLP            & $0.44 \pm .01$ & $0.77 \pm .03$ & $0.96 \pm .01$ & $0.97 \pm .04$ & $0.62 \pm .01$ & $0.51 \pm .01$ & 43 \\
SVM            & $0.32 \pm .00$ & $0.68 \pm .03$ & $0.91 \pm .00$ & ---            & ---            & $0.39 \pm .00$ & 832 \\
Image-CNN      & $0.30 \pm .03$ & $0.65 \pm .09$ & $0.87 \pm .04$ & $0.97 \pm .02$ & $0.51 \pm .06$ & $0.34 \pm .06$ & 31 \\
\midrule
\multicolumn{8}{l}{\textit{Temporal models}} \\
\textbf{LSTM}     & $0.75 \pm .02$ & $0.87 \pm .01$ & $0.999$        & $0.98 \pm .03$ & $0.86 \pm .02$ & $\mathbf{0.78 \pm .02}$ & 129 \\
CNN+Transformer   & $0.76 \pm .02$ & $0.82 \pm .03$ & $0.999$        & $0.99 \pm .00$ & $0.80 \pm .02$ & $0.78 \pm .01$ & 515 \\
GRU               & $0.74 \pm .04$ & $0.85 \pm .03$ & $0.998$        & $0.97 \pm .02$ & $0.83 \pm .04$ & $0.76 \pm .04$ & 119 \\
1D-CNN            & $0.73 \pm .04$ & $0.87 \pm .03$ & $0.998 \pm .001$ & $0.96 \pm .04$ & $0.82 \pm .03$ & $0.75 \pm .03$ & 77 \\
Transformer-small & $0.66 \pm .06$ & $0.75 \pm .07$ & $0.997 \pm .002$ & $0.99 \pm .01$ & $0.72 \pm .05$ & $0.69 \pm .06$ & 374 \\
\bottomrule
\end{tabular}
\end{table*}

\subsection{Comparison to Published CIC-IDS2017 Results}

Table~\ref{tab:priorwork} places our headline numbers alongside the most directly comparable Transformer-IDS studies.
Headline accuracy is not a useful comparison: CIC-IDS2017 is ${\sim}$80\% BENIGN, so predicting BENIGN throughout yields close to 80\% accuracy without learning anything useful.
Liu \emph{et al.}~\cite{liu2025cnnlstmtransformer} and Yao \emph{et al.}~\cite{yao2023cnntransformer} report accuracy and per-class precision/recall but never macro-F1; our Transformer ($99.89\%$) sits in the same range, yet macro-F1 tells a different story about rare-class detection that accuracy alone cannot.

Second, neither prior study supplies its temporal modules with a genuine flow sequence: Liu's implementation reshapes the feature map to $(B,1,128)$ before the LSTM and $(B,1,512)$ before the Transformer, while Yao operates on a single concatenated feature vector. Under a real $T{=}20$ sequence with leakage-free group evaluation, our Transformer drops to macro-F1 $0.55$ ($-0.23$ relative to random, almost entirely attributable to the padding protocol rather than the split itself), while a class-balanced Random Forest holds at $0.78$ ($+0.009$). Neither effect is detectable under the evaluation protocol adopted in prior work.

\begin{table*}[htb]
\centering
\caption{Comparison with prior Transformer-based IDS results on CIC-IDS2017. Values for prior work are taken from the respective papers; ``--'' indicates a metric not reported. $T$ denotes the sequence length seen by the recurrent or attention layer, not the CNN feature extractor. Our entries are mean over three seeds.}
\label{tab:priorwork}
\renewcommand{\arraystretch}{1.05}
\setlength{\tabcolsep}{4pt}
\footnotesize
\begin{tabular}{p{3.4cm}lcccccp{3.5cm}}
\toprule
\textbf{Paper / Setting} & \textbf{Architecture} & \textbf{$T$} & \textbf{Classes} & \textbf{Acc.} & \textbf{Macro-F1} & \textbf{Split} & \textbf{Notes} \\
\midrule
Liu \emph{et al.} 2025~\cite{liu2025cnnlstmtransformer}                  & CNN+LSTM+Transformer    & 1   & 9 (subset)    & 0.9920 & ---   & random           & Per-class supports inconsistent with published dataset; T=1 per Algorithm 1. \\
Yao \emph{et al.} 2023~\cite{yao2023cnntransformer}                      & CNN+Transformer         & 1   & 7 (collapsed) & 0.9106 & ---   & random/test      & Web attacks merged; Patator precision $0.28$, Infiltration DR $0.43$. \\
Yao \emph{et al.} 2023, 10-fold CV~\cite{yao2023cnntransformer}          & CNN+Transformer         & 1   & 7 (collapsed) & 0.9215 & ---   & 10-fold CV       & Same architecture as above.\\
\midrule
\textbf{Ours} (random, LSTM)                                              & LSTM over time          & 20  & 13            & 0.9985 & 0.776 & random           & Best macro-F1 of all 9 architectures; $+0.007$ over RF.\\
\textbf{Ours} (random, RF)                                                & Random Forest (last)    & 1   & 13            & 0.9978 & 0.769 & random           & Static baseline still in the top cluster. \\
\textbf{Ours} (random, Transformer)                                       & CNN+Transformer ([CLS]) & 20  & 13            & 0.9989 & 0.775 & random           & Highest accuracy; ties the LSTM on macro-F1.\\
\textbf{Ours} (full 15-class flow, RF)                                    & Random Forest           & 1   & 15            & 0.992  & 0.801 & random           & Best 15-class macro-F1; 5\,s training time.\\
\textbf{Ours} (group-split, RF)                                           & Random Forest (last)    & 1   & 13            & 0.9979 & 0.778 & group-by-5-tuple & Most robust model overall; $+0.009$ vs.\ random.\\
\textbf{Ours} (group-split, LSTM)                                         & LSTM over time          & 20  & 13            & 0.9964 & 0.753 & group-by-5-tuple & $-0.023$ vs.\ random.\\
\textbf{Ours} (group-split, GRU)                                          & GRU over time           & 20  & 13            & 0.9944 & 0.736 & group-by-5-tuple & $-0.027$ vs.\ random.\\
\textbf{Ours} (group-split, 1D-CNN)                                       & 1D-CNN over time        & 20  & 13            & 0.9969 & 0.733 & group-by-5-tuple & $-0.021$ vs.\ random.\\
\textbf{Ours} (group-split, Transformer)                                  & CNN+Transformer ([CLS]) & 20  & 13            & 0.9717 & 0.549 & group-by-5-tuple & $-0.226$ vs.\ random, of which $-0.243$ is the padding change (Table~\ref{tab:padding}).\\
\bottomrule
\end{tabular}
\end{table*}

Lightweight RNN/CNN models and a class-balanced Random Forest remain robust on a real temporal sequence assessed with macro-F1 under a leakage-free split, while the Transformer proves sensitive to padding convention---a sensitivity that the conventional protocol used in prior work cannot reveal.

\subsection{Sequence Length Ablation}
\label{sec:sweep}

\begin{figure}[htb]
\centering
\includegraphics[width=\columnwidth]{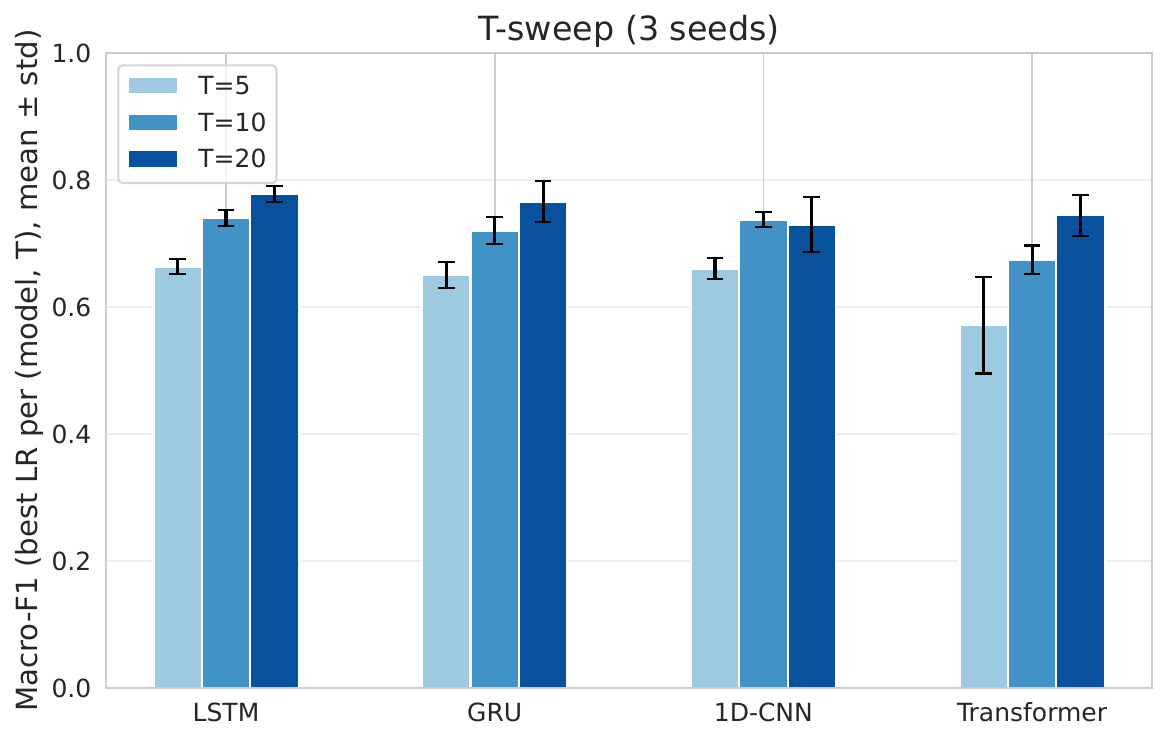}
\caption{Sequence length ablation across all four temporal models, $T \in \{5, 10, 20\}$. Each bar is the mean macro-F1 across three seeds at the learning rate with the best seed-mean for that ($T$, model) combination; error bars show $\pm 1$ std. $T{=}20$ is best or within one standard deviation of best for every model.}
\label{fig:tsweep-all}
\end{figure}

We swept $T \in \{5, 10, 20\}$ for each of the four temporal architectures, combined with a learning-rate search over $\{10^{-3}, 5{\cdot}10^{-4}\}$ for LSTM, GRU, and 1D-CNN, and over $\{10^{-3}, 5{\cdot}10^{-4}, 10^{-4}\}$ for
the Transformer, using three random seeds throughout. For each $(T, \text{model})$ combination we select the learning rate with the highest mean macro-F1 across seeds and report mean${}{\pm}{}$std at that rate; Table~\ref{tab:tsweep} and Fig.~\ref{fig:tsweep-all} present the results.

\begin{table}[!t]
\centering
\caption{Sequence-length ablation. Each cell reports the mean${}\pm{}$std macro-F1 across three seeds at the learning rate with the best seed-mean for that ($T$, model) combination, on the random 80/20 split with repeat-last padding. Best macro-F1 per row in \textbf{bold}; $T{=}20$ is best for three of the four models and within one standard deviation of best for the 1D-CNN.}
\label{tab:tsweep}
\renewcommand{\arraystretch}{1.05}
\footnotesize
\setlength{\tabcolsep}{3pt}
\begin{tabular}{lcccc}
\toprule
\textbf{Model} & $\boldsymbol{T{=}5}$ & $\boldsymbol{T{=}10}$ & $\boldsymbol{T{=}20}$ & $\boldsymbol{\Delta_{T=5\to20}}$ \\
\midrule
LSTM         & $0.664 \pm .011$ & $0.740 \pm .012$ & $\mathbf{0.778 \pm .012}$ & $+0.114$ \\
GRU          & $0.651 \pm .021$ & $0.720 \pm .021$ & $\mathbf{0.766 \pm .032}$ & $+0.115$ \\
1D-CNN       & $0.661 \pm .016$ & $\mathbf{0.738 \pm .012}$ & $0.730 \pm .043$ & $+0.069$ \\
Transformer  & $0.571 \pm .076$ & $0.674 \pm .022$ & $\mathbf{0.745 \pm .032}$ & $+0.173$ \\
\bottomrule
\end{tabular}
\end{table}

Three of the four temporal models improve monotonically with $T$ (the 1D-CNN peaks at $T{=}10$, with its $T{=}20$ score within one standard deviation of that peak); the gain from $T{=}5$ to the best horizon ranges from $+0.08$ for the 1D-CNN to $+0.17$ for the Transformer, which has the most to gain from a real temporal axis. $T{=}20$ is best or statistically indistinguishable from best for every model, so we adopt it as the fixed horizon for the remainder of the paper. We did not sweep beyond $T{=}20$ for two reasons: first, CIC-IDS2017's short conversation lengths mean that larger windows increase padding rather than genuine temporal signal (at $T{=}20$, approximately 73\% of windows already require padding); second, the Transformer's inference latency grows steeply with $T$ and a full three-seed sweep at larger horizons would incur substantial computational cost without a clear benefit. At $T{=}20$ the four temporal architectures are within $0.048$ macro-F1 of each other.

\subsection{Inference Latency and Cost--Accuracy Trade-off}
\label{sec:latency}

\begin{figure}[!t]
\centering
\includegraphics[width=\columnwidth]{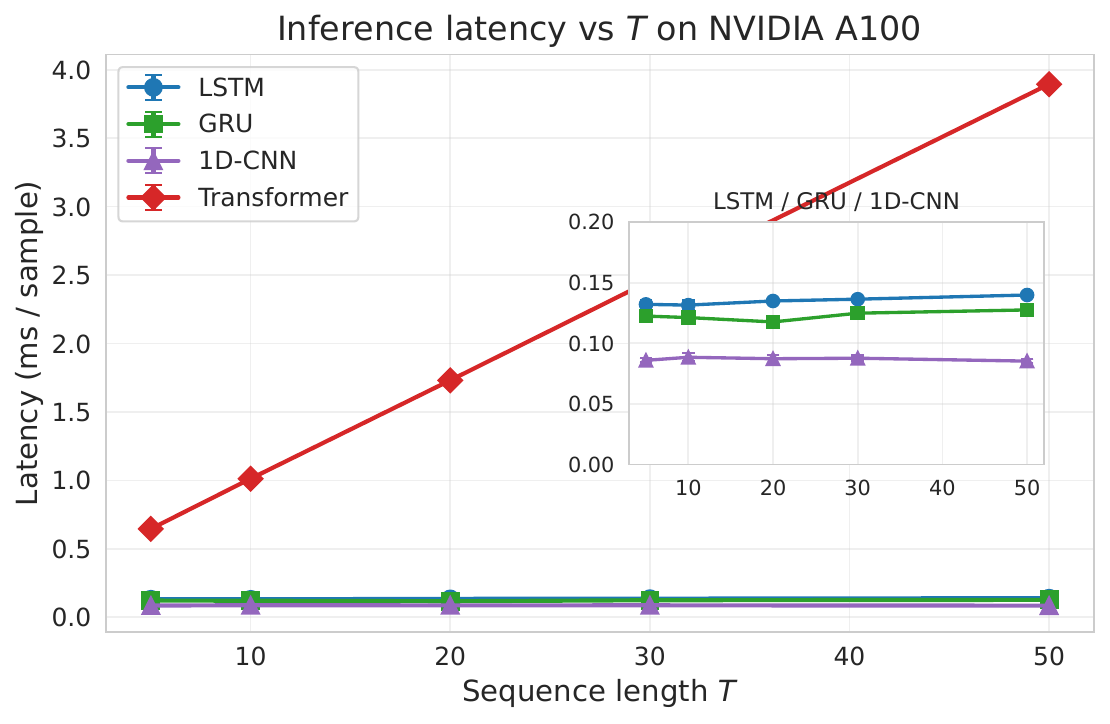}
\caption{Per-sample inference latency vs.\ sequence length $T$ on an NVIDIA A100 GPU, mean${}{\pm}{}$std over 20 repetitions of a 128-window batch. Main panel: all four temporal models (Transformer in red; LSTM/GRU/1D-CNN clustered near zero); inset: zoom below $0.2$\,ms. The Transformer grows ${\sim}6{\times}$ from $T{=}5$ to $T{=}50$; constant overheads dominate at these sequence lengths so the $\mathcal{O}(T^{2})$ self-attention term is not yet apparent. LSTM, GRU, and 1D-CNN remain near-constant.}
\label{fig:latency}
\end{figure}

Fig.~\ref{fig:latency} reports per-sample forward-pass latency on an A100 GPU as $T$ varies (twenty repetitions of a 128-window batch per ($T$, model)). The CNN--Transformer's cost grows steeply and approximately linearly over the measured range, from $0.65$\,ms at $T{=}5$ to $3.89$\,ms at $T{=}50$ (the quadratic self-attention term is not yet dominant at these sequence lengths). LSTM, GRU, and 1D-CNN all stay below $0.15$\,ms across the same range and remain near-constant. At $T{=}20$, the Transformer's inference latency ($1.73$\,ms) is already an order of magnitude greater than the LSTM's ($0.13$\,ms), the GRU's ($0.12$\,ms), and the 1D-CNN's ($0.09$\,ms). Random Forest inference on the same batch measures $0.27$\,ms/sample on CPU. 

\begin{figure}[!t]
\centering
\includegraphics[width=\columnwidth]{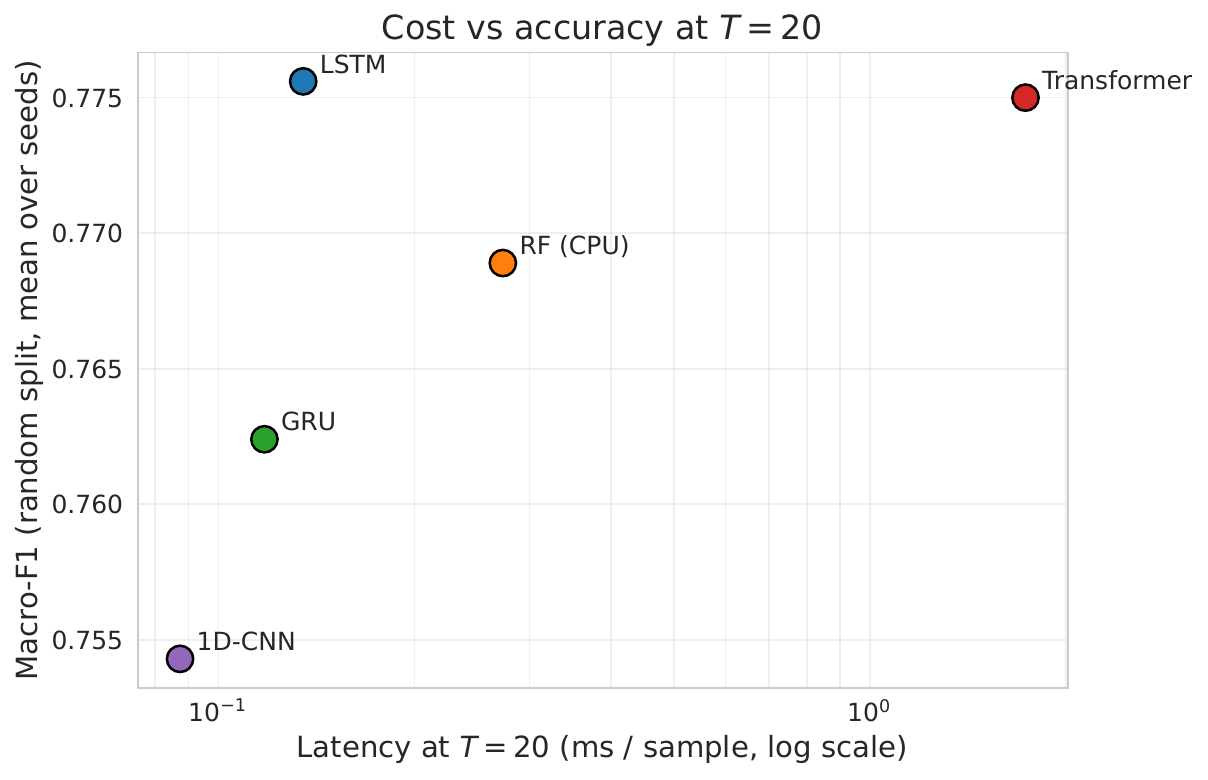}
\caption{Cost vs.\ accuracy at $T{=}20$ (log-scale latency, mean macro-F1 over three seeds; RF latency on CPU). The 1D-CNN, GRU and LSTM form the Pareto frontier; the CNN--Transformer matches the LSTM's macro-F1 at ${\sim}13\times$ the latency, and the Random Forest is likewise dominated.}
\label{fig:pareto}
\end{figure}

Combining latency with random-split macro-F1(Fig.~\ref{fig:pareto}), the LSTM Pareto-dominates the CNN--Transformer at $T{=}20$: an equal macro-F1 ($0.776$ vs.\ $0.775$) at $0.13$ vs.\ $1.73$\,ms per sample. The 1D-CNN is the fastest model overall ($0.09$\,ms), trading $0.02$ macro-F1 for ${\sim}20\times$ lower latency than the Transformer.

\subsection{Ablation Studies}
\label{sec:ablations}

Three follow-up experiments test how robust the Table~\ref{tab:ids_splits} numbers really are.

\subsubsection{Static Models on the Full 15-Class Flow Set}
\label{sec:static15}

The windowing constraint forces us to drop two classes (Heartbleed, DoS Slowhttptest) that cannot produce stratifiable windows. To check whether this disadvantages the static baselines unfairly, we re-trained SVM, RF, MLP, and image-CNN on the original \emph{flow-level} CIC-IDS2017 data with all 15 classes and a per-class cap of $25{,}000$ to balance training. The cap never sub-samples the rare classes --- every rare-class flow that falls in the training split is kept (e.g.\ ${\sim}9$ of the 11 Heartbleed flows in expectation) --- and only reduces the dominant ones.

\begin{table}[!t]
\centering
\caption{Static models on the full 15-class flow data with per-class cap of $25{,}000$ (no temporal windowing), mean${}\pm{}$std over three seeds. $\Delta$ is vs.\ the best temporal model in Table~\ref{tab:ids_splits} (LSTM, $0.776$).}
\label{tab:static15}
\renewcommand{\arraystretch}{1.05}
\footnotesize
\setlength{\tabcolsep}{2pt}
\begin{tabular}{lcccc}
\toprule
\textbf{Model} & \textbf{Acc.} & \textbf{F1$_{\text{macro}}$} & $\boldsymbol{\Delta_{\text{F1}}}$ & \textbf{Train (s)} \\
\midrule
\textbf{Random Forest-15}    & $0.992 \pm .001$ & $\mathbf{0.801 \pm .030}$ & \hspace{2mm}$+0.025\,\up$ & 4.8 \\
MLP-15                       & $0.871 \pm .012$ & $0.487 \pm .011$          & $-0.289 \,\down$ & 30.5 \\
CNN-image-15                 & $0.875 \pm .012$ & $0.477 \pm .031$          & $-0.299 \,\down$ & 25.5 \\
SVM-15                       & $0.810 \pm .022$ & $0.363 \pm .023$          & $-0.413 \,\down$ & 1605.8 \\
\bottomrule
\end{tabular}
\end{table}

The Random Forest-15 reaches macro-F1 $\mathbf{0.801 \pm .030}$ (Table~\ref{tab:static15}), narrowly above the best temporal model on the windowed 13-class subset (LSTM, $0.78$) and detecting all three additional classes (Heartbleed, DoS Slowhttptest, Infiltration) that the windowed pipeline either drops or fails to detect at the flow level. Whether the cleaner $13$-class macro-F1 metric or the harder full-$15$-class metric is the right comparison is itself an evaluation choice; the fact that a Random Forest that trains in under five seconds is competitive on either bar is the point.

\subsubsection{Padding Scheme: Zero-pad+Mask vs.\ Repeat-last}
\label{sec:padding}

The primary protocol uses repeat-last padding to match prior IDS work. The cleaner alternative is to fill short conversations with zeros and pass an explicit per-position mask to the model. We re-ran all four temporal models, plus Transformer-small, with zero-pad+mask, three seeds each.

\begin{table}[!t]
\centering
\caption{Macro-F1 under repeat-last padding (primary protocol) versus zero-padding plus an explicit attention mask, mean${}\pm{}$std over three seeds. $\Delta$ is the change when the repeat-last cue is removed.}
\label{tab:padding}
\renewcommand{\arraystretch}{1.05}
\footnotesize
\setlength{\tabcolsep}{4pt}
\begin{tabular}{lccc}
\toprule
\textbf{Model} & \textbf{Repeat-last} & \textbf{Zero-pad+mask} & $\boldsymbol{\Delta_{\text{F1}}}$ \\
\midrule
LSTM              & $0.78 \pm .02$ & $0.76 \pm .01$ & $-0.017 \,\down$ \\
GRU               & $0.76 \pm .04$ & $0.74 \pm .01$ & $-0.019 \,\down$ \\
1D-CNN            & $0.75 \pm .03$ & $0.71 \pm .03$ & $-0.046 \,\down$   \\
Transformer       & $0.78 \pm .01$ & $0.53 \pm .03$ & $\mathbf{-0.243} \,\down$ \\
Transformer-small & $0.69 \pm .06$ & $0.40 \pm .08$ & $\mathbf{-0.289} \,\down$ \\
\bottomrule
\end{tabular}
\end{table}

Repeat-last padding sticks the same final flow vector into every padded position, so self-attention can use that repetition as a (very strong) feature. Zero-pad with an attention mask blocks the model from attending to those positions at all, so it has to learn from the genuine flows alone. Recurrent and convolutional inductive biases are largely invariant to the choice (the LSTM and GRU lose ${\sim}\,0.02$ and the 1D-CNN $0.05$). The two Transformer variants depend strongly on the repeat-last cue: removing it costs the full Transformer $0.24$ macro-F1 and Transformer-small $0.29$, with Transformer-small showing substantially higher per-seed variance ($\pm 0.08$) than any other model in the experiment, indicating the zero-pad+mask setting is harder for self-attention to optimise consistently.

\emph{Masking implementation note.} LSTM and GRU receive a Keras \texttt{Masking(mask\_value=0.0)} layer that gates gradient flow at padded positions. The 1D-CNN receives only zeroed inputs without a gating layer, because \texttt{Conv1D} silently discards Keras masks; this difference in mechanism may partly explain its distinct behaviour ($-0.046$) relative to the recurrent models (${\approx}{-}0.02$).

\subsubsection{Non-Padded Subset}
\label{sec:nonpad}

A complementary check restricts evaluation to the subset of windows containing no padding. This subset accounts for ${\sim}27\%$ of windows and covers 9 classes; Bot, FTP-Patator, Infiltration, and SSH-Patator are dropped because their five-tuple conversations rarely contain $T{=}20$ real consecutive flows.

\begin{table}[!t]
\centering
\caption{Macro-F1 on the full padded set vs.\ the non-padded subset (${\sim}27\%$ of windows, 9 classes), mean${}\pm{}$std over three seeds. \emph{Every} temporal model improves on the non-padded subset.}
\label{tab:nopad}
\renewcommand{\arraystretch}{1.05}
\footnotesize
\setlength{\tabcolsep}{4pt}
\begin{tabular}{lccc}
\toprule
\textbf{Model} & \textbf{Padded set} & \textbf{Non-padded} & $\boldsymbol{\Delta_{\text{F1}}}$ \\
\midrule
LSTM              & $0.78 \pm .02$ & $0.85 \pm .01$ & $+0.076\,\up$ \\
GRU               & $0.76 \pm .04$ & $0.86 \pm .01$ & $+0.093\,\up$ \\
1D-CNN            & $0.75 \pm .03$ & $0.82 \pm .02$ & $+0.067\,\up$ \\
\textbf{Transformer} & $0.78 \pm .01$ & $\mathbf{0.89 \pm .02}$ & $+0.118\,\up$ \\
Transformer-small & $0.69 \pm .06$ & $0.83 \pm .04$ & $+0.141\,\up$ \\
\bottomrule
\end{tabular}
\end{table}

\emph{Every} temporal architecture's macro-F1 \emph{rises} when evaluation is restricted to genuinely sequential windows (gains of $0.07$--$0.14$). This indicates the padded fraction of the full set --- whether under repeat-last or zero-pad+mask --- is harder than the underlying real-sequence problem for all temporal models, not just the Transformer. Combined with Table~\ref{tab:padding}, the picture is consistent: the LSTM, GRU and 1D-CNN are robust to the choice of padding scheme and improve on real sequences; the Transformer achieves the best macro-F1 of the entire experiment on
genuinely sequential input ($0.89$) yet is fragile to the loss of the repeat-last cue under masked attention.

\section{Realistic Evaluation}\label{sec:realistic}

We now evaluate all nine architectures under the two leakage-free protocols defined in Section~\ref{sec:splits}. Both protocols use zero-pad+mask windows; the padding contribution to any observed drop is isolated by the matched baseline in Section~\ref{sec:padding}.

\subsection{Time-Based Split (Degenerate)}
\label{sec:realistic-time}

We sort all $2.83$\,M flows by timestamp and use the chronological first 80\% as training and last 20\% as test. The result is degenerate on this dataset: only two of the 13 classes survive in the test set, because CIC-IDS2017's attacks were scheduled day by day~\cite{sharafaldin2018cicids}. Most attack labels appear only in the chronological training window. Every model collapses to or near the majority-class baseline (macro-F1 $\approx 0.50$): the LSTM and GRU reach $0.52$, and every other model sits between $0.48$ and $0.50$. We treat this as a property of CIC-IDS2017's attack scheduling rather than a useful generalisation benchmark.

\subsection{Group-by-Five-Tuple Split}
\label{sec:realistic-group}

The cleaner generalisation test partitions five-tuple groups, not flows. Each five-tuple is assigned to either train or test (80/20), so every window from a given conversation lands on one side. This produces ${\sim}503\,$k training and ${\sim}127\,$k test windows across all 13 classes. Models can no longer memorise specific conversations; they have to generalise to conversations they have never seen.

\begin{table}[!t]
\centering
\caption{Macro-F1 under the leakage-free group-by-five-tuple split, mean${}\pm{}$std over three seeds, sorted by macro-F1. $\Delta$ is vs.\ the random-split macro-F1 in Table~\ref{tab:ids_splits}.}
\label{tab:group}
\renewcommand{\arraystretch}{1.05}
\footnotesize
\setlength{\tabcolsep}{4pt}
\begin{tabular}{lccc}
\toprule
\textbf{Model} & \textbf{Random} & \textbf{Group} & $\boldsymbol{\Delta_{\text{F1}}}$ \\
\midrule
\textbf{RF}       & $0.77 \pm .01$ & $\mathbf{0.78 \pm .04}$ & \hspace{2mm}$+0.009\,\up$   \\
LSTM              & $0.78 \pm .02$ & $0.75 \pm .03$ & $-0.023 \,\down$ \\
GRU               & $0.76 \pm .04$ & $0.74 \pm .05$ & $-0.027 \,\down$ \\
1D-CNN            & $0.75 \pm .03$ & $0.73 \pm .01$ & $-0.021 \,\down$ \\
CNN+Transformer   & $0.78 \pm .01$ & $0.55 \pm .09$ & $-0.226 \,\down$ \\
MLP               & $0.51 \pm .01$ & $0.52 \pm .01$ & \hspace{2mm}$+0.019\,\up$   \\
Image-CNN         & $0.34 \pm .06$ & $0.41 \pm .01$ & \hspace{2mm}$+0.062\,\up$   \\
SVM               & $0.39 \pm .00$ & $0.40 \pm .02$ & \hspace{2mm}$+0.008\,\up$   \\
Transformer-small & $0.69 \pm .06$ & $0.38 \pm .03$ & \hspace{-0.85mm}$\mathbf{-0.303} \,\down$ \\
\bottomrule
\end{tabular}
\end{table}

\begin{figure*}[!t]
\centering
\begin{subfigure}{0.49\textwidth}
\centering
\includegraphics[width=\linewidth]{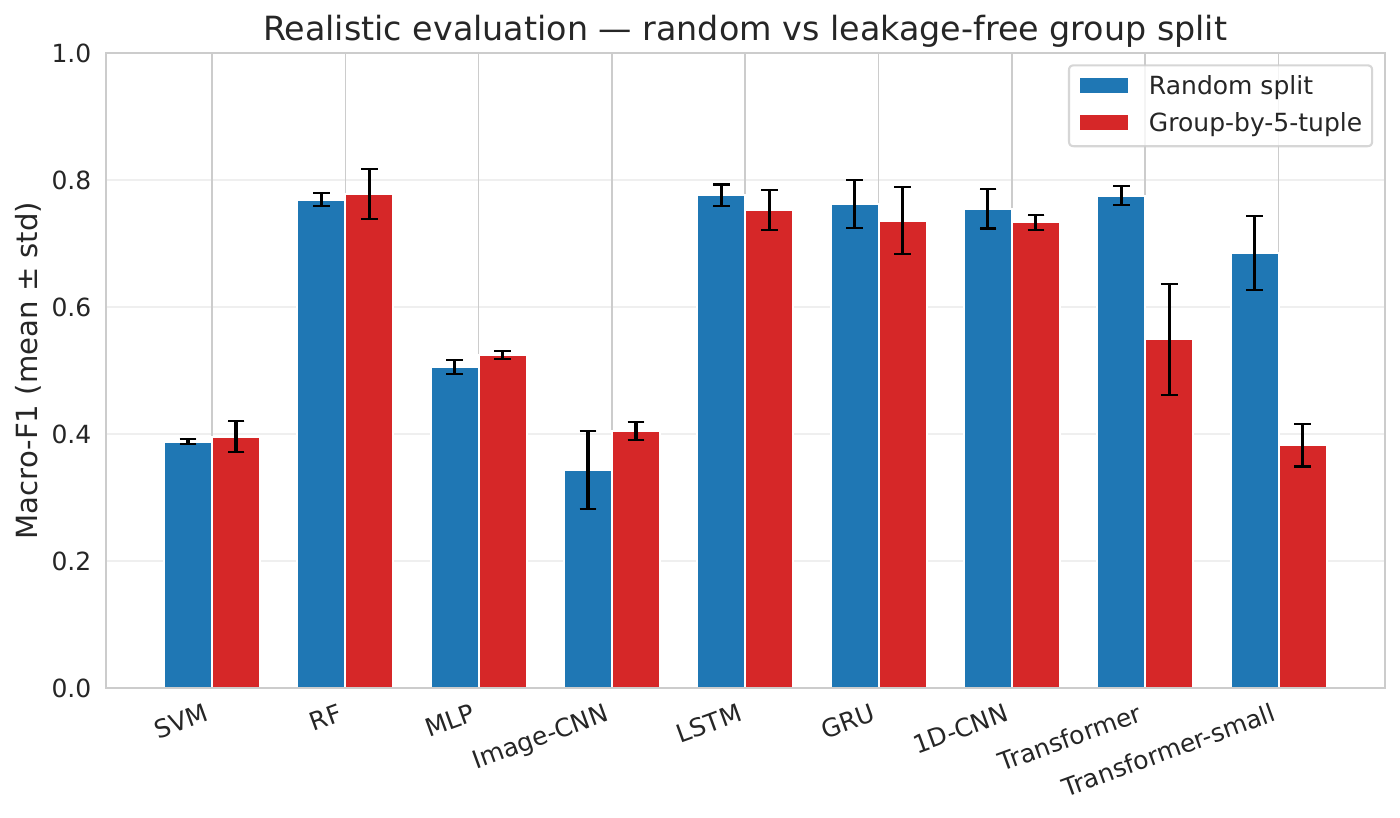}
\caption{Macro-F1 under each split.}
\label{fig:realistic}
\end{subfigure}
\hfill
\begin{subfigure}{0.49\textwidth}
\centering
\includegraphics[width=\linewidth]{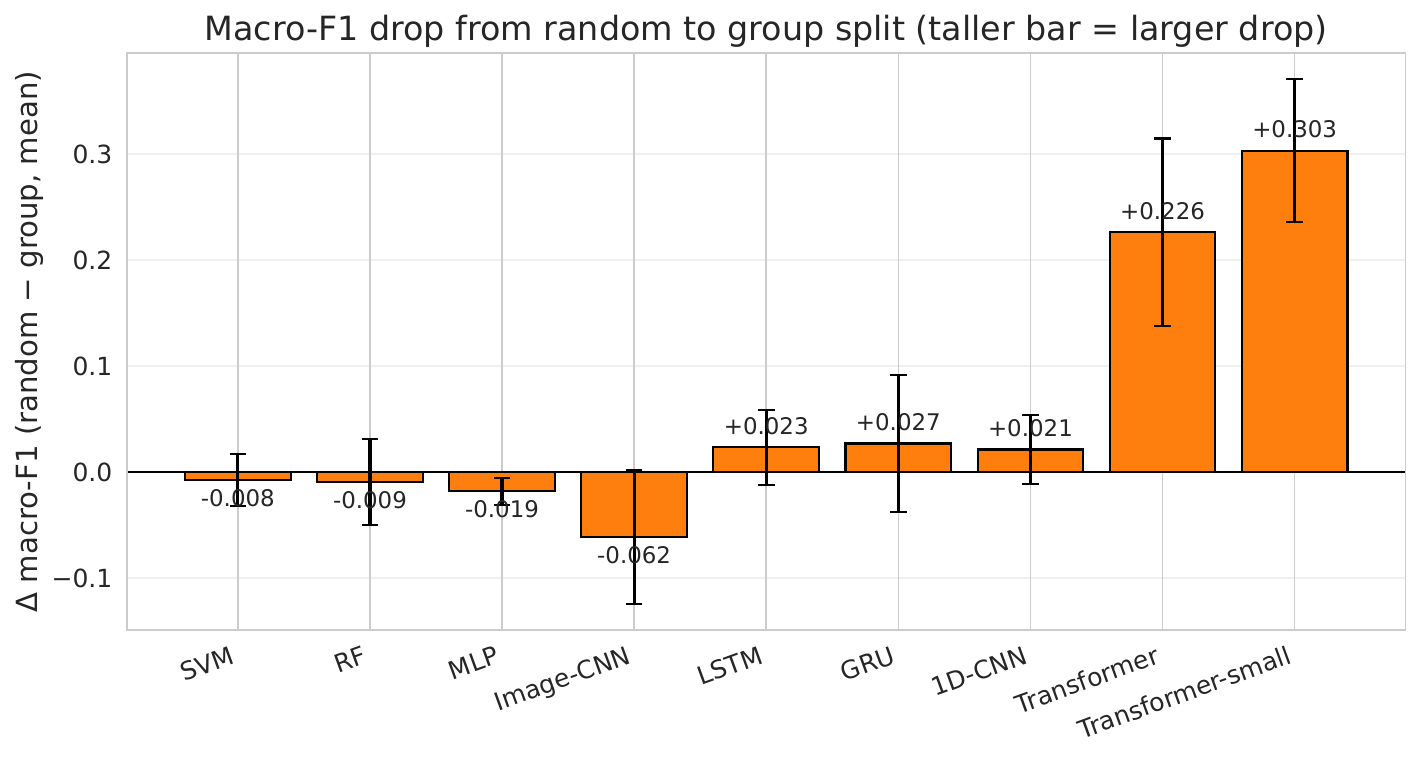}
\caption{Macro-F1 drop from random.}
\label{fig:realistic-drops}
\end{subfigure}
\caption{Realistic evaluation, mean${}\pm{}$std over three seeds. Left: macro-F1 under random vs.\ group split. Right: drop from the random split per model. The Random Forest and the lightweight RNN/CNN models hold within $0.03$ of their random-split macro-F1; the CNN--Transformer drops $0.23$, almost entirely attributable to the zero-pad+mask protocol rather than the split itself (cf.\ Table~\ref{tab:padding}).}
\end{figure*}

\begin{figure*}[!t]
\centering
\includegraphics[width=\textwidth]{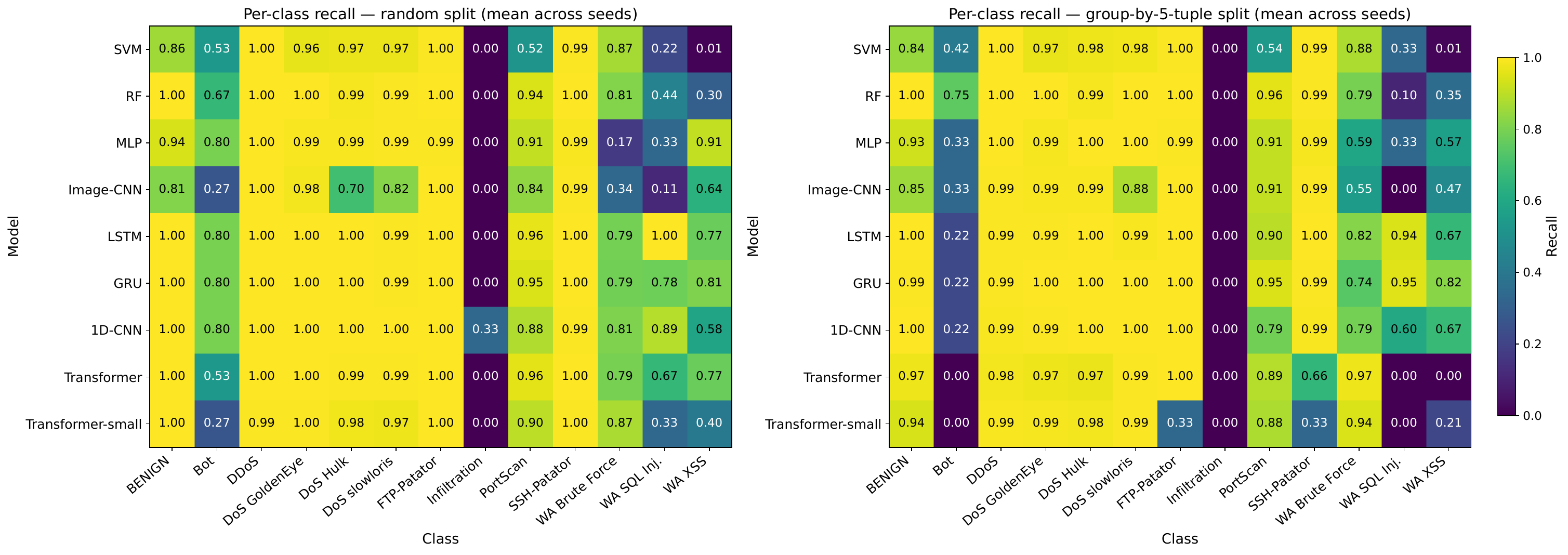}
\caption{Per-class recall under the random split (left) vs.\ the leakage-free group-by-five-tuple split (right), mean across three seeds. The Random Forest and SVM (top rows) hold the high-volume classes under both splits; the lightweight temporal models lose recall mainly on Bot, while the two Transformer variants also lose rare web-attack and Patator classes.}
\label{fig:perclass-by-split}
\end{figure*}

\begin{figure}[!t]
\centering
\includegraphics[width=\columnwidth]{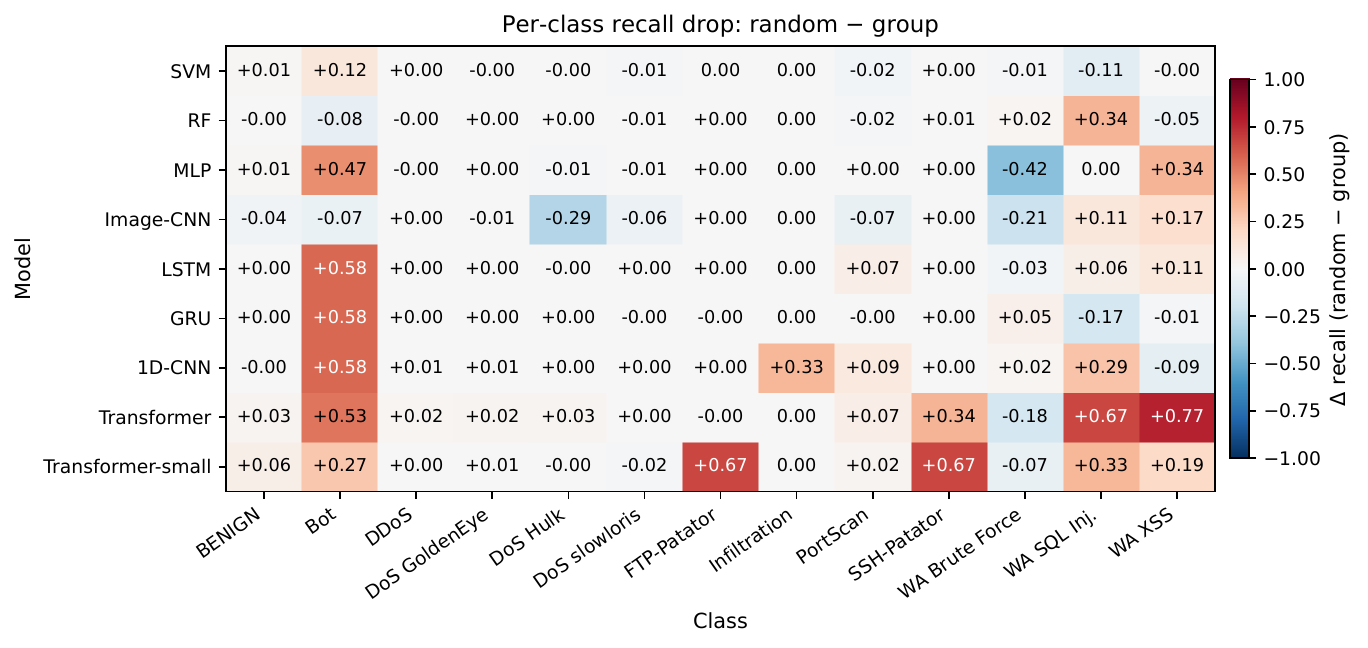}
\caption{Per-class recall drop (random $-$ group), mean across three seeds. Red cells: recall loss going to the leakage-free split; blue cells: gain. The Bot column drops for every temporal architecture (the Random Forest's Bot recall actually improves); the two Transformer variants lose recall across more classes than the recurrent and convolutional models do.}
\label{fig:perclass-drop}
\end{figure}

Table~\ref{tab:group} and Figs.~\ref{fig:realistic}--\ref{fig:realistic-drops} present the macro-F1 results; per-class recall by split and the recall drop are shown in Figs.~\ref{fig:perclass-by-split} and~\ref{fig:perclass-drop}. The class-balanced Random Forest is the most robust model in the experiment: its macro-F1 \emph{rises} slightly under the group split ($0.769 \to 0.778$). The lightweight RNN/CNN models (LSTM, GRU, 1D-CNN) cluster just below at macro-F1 $0.73$--$0.75$ (deltas of $-0.023$, $-0.027$, $-0.021$ respectively, all within one per-seed standard deviation). The CNN--Transformer drops by $0.23$ to $0.55$, but comparison against the matched zero-pad baseline (Table~\ref{tab:padding}) shows the split itself costs it almost nothing ($0.53$ random vs.\ $0.55$ group): the drop is attributable to the padding protocol, not to conversation memorisation. Transformer-small behaves the same way ($-0.303$ vs.\ random, $-0.014$ vs.\ the matched baseline). The remaining static baselines (MLP, image-CNN, SVM) improve modestly under the group split but stay weak under both.

Per-class recall (Figs.~\ref{fig:perclass-by-split},~\ref{fig:perclass-drop}) sharpens the structural reading. The high-volume classes are essentially unaffected: DDoS, FTP-Patator and SSH-Patator are recalled at ${\geq}0.99$ by nearly every model under both splits, and the Random Forest holds every class within $0.02$ of its random-split recall except Web Attack--SQL Injection (a 16-window class). The losses concentrate on the rare classes and on self-attention: every temporal model loses about $0.5$--$0.6$ recall on Bot (LSTM, GRU and 1D-CNN all drop $0.80 \to 0.22$), while the Transformer additionally collapses on Web Attack--XSS ($0.77 \to 0.00$), Web Attack--SQL Injection ($0.67 \to 0.00$) and SSH-Patator ($1.00 \to 0.66$), and Transformer-small loses FTP-Patator and SSH-Patator ($1.00 \to 0.33$ each). The reading is structural: rare-class recall on unseen conversations depends on inductive bias, and recurrent, convolutional and tree-based models transfer better than self-attention on this benchmark. Operationally, the Random Forest's false-alarm rate is essentially unchanged between splits ($0.16\% \to 0.14\%$), while the Transformer's grows by an order of magnitude ($0.04\% \to 2.7\%$).

\section{Discussion}\label{sec:discussion}

Taken together, the ablations (Section~\ref{sec:ablations}) and the leakage-free evaluation (Section~\ref{sec:realistic}) point to a single underlying explanation: the benefit of any model on CIC-IDS2017 depends on the alignment between its inductive bias, its input representation, and the evaluation protocol---not on architectural sophistication.

\textbf{Recurrent and convolutional inductive biases generalise well to unseen conversations.} LSTM, GRU, and 1D-CNN show consistent behaviour across every condition tested: stable under the group split, stable under the padding ablation, and improved on the non-padded subset. This consistency across three independent tests is the strongest evidence that their random-split numbers reflect genuine sequence learning rather than an artefact of any single evaluation choice.

\textbf{The Transformer is not architecturally weak; it is protocol-dependent.} Its best result overall (macro-F1 $0.89$) occurs precisely when it receives the input it was designed for---a complete, unpadded sequence. The same model also produces the largest false-alarm increase observed once that condition is violated. This is a sharper and more useful finding than ``Transformers underperform'': it identifies padding convention, not self-attention itself, as the variable that determines whether the architecture succeeds or fails.

\textbf{The Random Forest's robustness is a consequence of the windowing pipeline, not of its decision-tree structure per se.} Its stability under the group split is consistent with reduced reliance on dominant-class statistics following the increased subsample diversity adopted here (Section~\ref{sec:methodology}), and with the destination-port-shortcut overfitting documented by Engelen \emph{et al.}~\cite{engelen2021troubleshoot} no longer dominating its decisions. This suggests RF robustness on CIC-IDS2017 is sensitive to subsampling strategy and should not be read as a general property of tree ensembles on IDS data.

\textbf{Sequence length and padding cleanliness are the two levers that matter most for temporal models.} The $T$-sweep (Section~\ref{sec:sweep}) and the non-padded subset (Section~\ref{sec:nonpad}) isolate these levers independently, and both point in the same direction: every temporal architecture benefits from longer, cleaner sequences, with the Transformer gaining the most from each. This is consistent with self-attention having the greatest capacity to exploit additional genuine temporal structure, and the least tolerance for synthetic structure introduced by padding.

\textbf{Architectural asymmetry.} The CNN--Transformer pre-encodes each flow with a CNN before the Transformer encoder, while LSTM and GRU operate on raw $z$-score normalised features. The Transformer's padding sensitivity may partly reflect this input-representation difference rather than self-attention alone; a Transformer operating directly on raw features, without a CNN pre-encoder, would help isolate the two effects, though it would not address the architecture's underlying latency cost, which is driven by self-attention's quadratic complexity in $T$ (Section~\ref{sec:latency}) rather than by the choice of pre-encoder. Reducing that cost would instead require efficient attention variants (e.g.\ linear or sparse attention mechanisms) that scale sub-quadratically with sequence length; whether such variants retain the Transformer's strong performance on clean sequences while improving robustness to padding is an open question we leave for future work.

\textbf{Limitations.} Results are specific to CIC-IDS2017; generalisation to datasets with longer sessions or different attack distributions (e.g.\ UNSW-NB15) is not established. All latency measurements are on an NVIDIA A100 GPU; CPU-only inference relevant to edge IDS deployments may change relative orderings. Macro-F1 is used as the headline metric; the FPR analysis provides complementary operational evidence but is not exhaustive.

\section{Conclusion}\label{sec:conclusion}

We reformulated CIC-IDS2017 as a genuine temporal sequence-classification task and benchmarked nine architectures across four evaluation conditions designed to reflect realistic intrusion-detection requirements.
Across these conditions, padding convention and split protocol, not architecture, governed reported performance: the Transformer was the strongest sequence modeller given clean, genuinely sequential input, yet alone among the models it collapsed once the repeat-last padding cue was removed, while lightweight recurrent, convolutional, and tree-based models remained stable.
Widely used random splits with repeat-last padding conceal both this fragility and the accompanying surge in false alarms entirely.

From an operational standpoint, the LSTM matches the Transformer on macro-F1 ($0.776$ vs.\ $0.775$) at $13{\times}$ lower inference latency, and the Random Forest is the most deployment-robust model, with stable macro-F1 and false-alarm rate across all evaluation conditions.
These results suggest that architectural sophistication alone is a poor predictor of real-world IDS performance, and that evaluation discipline deserves at least as much attention as model design in future work.
Practitioners deploying IDS on short-session datasets should treat the Transformer with caution unless clean-sequence conditions can be guaranteed.

Future work will extend this framework to additional datasets (UNSW-NB15, CIC-IoT-2023), Transformer variants without CNN pre-encoders, and online streaming scenarios with explicit false-positive-rate constraints.

\section*{Acknowledgment}
The experimental design, implementation, and manuscript preparation were carried out primarily by 2Lt~Z.\ Moczkodan. The research direction, technical review, and overall supervision were provided by Dr.\ H.\ Ragab. Early work on the
static modelling baselines and initial dataset exploration was conducted with the support of 2Lt~P.\ Parikh and 2Lt~A.\ Ben Rquia, under the co-supervision of Capt.\ J.\ Weibe (Cyber Training Unit, Canadian Armed Forces Cyber Command,
CAFCYBERCOM) and Dr.\ H.\ Ragab.

\bibliographystyle{IEEEtran}
\bibliography{references}

\begin{IEEEbiography}
[{\includegraphics[width=1in,height=1.25in,clip,keepaspectratio]{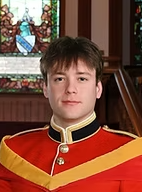}}]{Zach Moczkodan}(Student Member, IEEE) received the B.Eng. degree in Computer Engineering from the Royal Military College of Canada, Kingston, ON, Canada, in 2026. He is currently a Second Lieutenant in the Canadian Armed Forces. He was a recipient of the Prix Chikhani, and his capstone project was recognized in the IEEE Best Capstone Project competition. His research interests lie at the intersection of applied artificial intelligence and cybersecurity, with a focus on machine learning--based intrusion detection and network traffic analysis.
\end{IEEEbiography}

\begin{IEEEbiography}[{\includegraphics[width=1in,height=1.25in,clip,keepaspectratio]{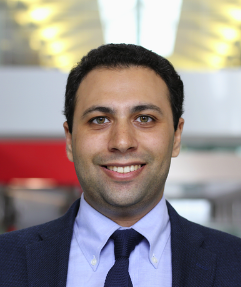}}]{Hany Ragab}

(Member, IEEE) received the Ph.D. degree in Electrical and Computer Engineering from Queen's University, Kingston, ON, Canada, in 2024. He is currently an Assistant Professor with the Department of Electrical and Computer Engineering, Royal Military College of Canada, Kingston, ON, Canada. His research interests include AI-enabled security and resilience of cyber-physical systems, with emphasis on anomaly and intrusion detection and trustworthy autonomy for intelligent transportation and defence. Dr. Ragab serves as a reviewer for several IEEE journals and international conferences and is a professional member of the Institute of Navigation (ION) and SAE International.
\end{IEEEbiography}

\end{document}